\documentclass[journal,compsoc]{IEEEtran}

\usepackage{color}
\usepackage{graphicx,epstopdf}
\usepackage{array}
\usepackage{url}
\usepackage{cite}
\usepackage[utf8]{inputenc}
\usepackage{ragged2e}
\usepackage{amsmath}
\usepackage{amssymb}
\usepackage{algorithm}
\usepackage{algpseudocode}
\usepackage{enumitem}
\usepackage{makecell}
\DeclareMathOperator*{\argmax}{argmax}

\fontfamily{pcr}\selectfont

\begin{document}

\title{Mobility Management in 5G and Beyond: A Novel Smart Handover with Adaptive Time-to-Trigger and Hysteresis Margin}

\author{Raja~Karmakar,~\IEEEmembership{Member,~IEEE},
     Georges~Kaddoum,~\IEEEmembership{Senior~Member,~IEEE}
     and~Samiran~Chattopadhyay,~\IEEEmembership{Senior~Member,~IEEE}

\thanks{R. Karmakar is with Department of Computer Science and Engineering, Techno International New Town, Kolkata, India 700156 (Email: raja.karmakar@tint.edu.in).}
\thanks{G. Kaddoum is with Department of Electrical Engineering, ETS, University of Quebec, Montreal, Canada (Email: georges.kaddoum@etsmtl.ca).}
\thanks{S. Chattopadhyay is with Department of Information Technology, Jadavpur University, Kolkata, India 700098, Institute for Advancing Intelligence, TCG Centres for Research and Education in Science and Technology, Kolkata, India (Email: samiran.chattopadhyay@jadavpuruniversity.in).}

} 

%\maketitle

\IEEEtitleabstractindextext{
\begin{abstract}

%With the fast increase in the volume of mobile users, next generation cellular networks are evolving to provide extremely high data rates that can support new applications demanding very high throughput and low delay. 
%In this direction, 
%Next generation cellular networks are evolving to support new applications demanding very high throughput and low delay. In this direction, 
The 5th Generation (5G) New Radio (NR) and beyond technologies will support enhanced mobile broadband, very low latency communications, and huge numbers of mobile devices. Therefore, for very high speed users, seamless mobility needs to be maintained during the migration from one cell to another in the handover. Due to the presence of a massive number of mobile devices, the management of the high mobility of a dense network becomes crucial. Moreover, a dynamic adaptation is required for the Time-to-Trigger (TTT) and hysteresis margin, which significantly impact the handover latency and overall throughput. Therefore, in this paper, we propose an online learning-based mechanism, known as {\em \textbf{L}earning-based \textbf{I}ntelligent \textbf{M}obility \textbf{M}anagement (LIM2)}, for mobility management in 5G and beyond, with an intelligent adaptation of the TTT and hysteresis values. LIM2 uses a Kalman filter to predict the future signal quality of the serving and neighbor cells, selects the target cell for the handover using {\em state-action-reward-state-action (SARSA)}-based reinforcement learning, and adapts the TTT and hysteresis using the {\em $\epsilon$-greedy} policy. %To thoroughly evaluate the performance of the LIM2 algorithm, 
We implement a prototype of the LIM2 in NS-3 and extensively analyze its performance, where it is observed that the LIM2 algorithm can significantly improve the handover operation in very high speed mobility scenarios. 

\end{abstract}
}

\maketitle

\IEEEdisplaynontitleabstractindextext

\IEEEpeerreviewmaketitle

\section{Introduction}
\label{sec:intro}

The 5th Generation (5G) New Radio (NR) standard is designed to support high data rates, extremely low latency (suitable for real-time applications), very high mobility of User Equipments (UEs), and higher energy efficiency. It is expected that 5G will provide $1000$ times higher data traffic volumes than current cellular networks, $100$ times more mobile connections, a peak data rate of $10$ Gbps, and more than $10$ Mbps per-user data rates~\cite{chih2016new,iwamura2015ngmn,agyapong2014design}. Moreover, the Internet of Things (IoT) and mobile Internet are considered as the two primary drivers of 5G mobile communication systems, and will cover a broad prospect for 5G and beyond due to their wide range of application perspective~\cite{akpakwu2017survey,chih2016new}.
For instance, a continuous network coverage should be maintained in high speed trains~\cite{3GPP5G,li2020beyond}. 
To provide {\em always-on} Internet access services, cellular communications, such as 5G and beyond, is a promising solution. %Since the mobile network is the largest wireless infrastructure, it can offer the mobility management for a wide area network access. The high speed mobile network has served billions of mobile users, and it is expected that it will serve trillions of IoT devices~\cite{li2020beyond,akpakwu2017survey}. As the current broadband cellular network technology is moving to 5G and beyond, there will be a tremendous increase in the density of connected devices in 5G networks that include high mobility UEs. As an integral part, high mobility requirements should be incorporated in 5G. For instance, a continuous network coverage should be maintained in high speed trains~\cite{3GPP5G,li2020beyond}. 

In cellular networks, the basic requirement from the mobility management entity is an efficient handover operation that can be executed without interruption~\cite{gures2020comprehensive,morales2015taming}. A {\em handover} is a mechanism in mobile communications, in which an ongoing call or a data access period is transferred from one gNodeB (gNB) (in 5G NR, the base station is known as gNB) to another one without disconnecting the current session. Therefore, when a UE is active (either in call or data session), the gNB applies an active signaling for the handover with a configurable link monitoring mechanism~\cite{yajnanarayana20205g,gures2020comprehensive}. Given the significantly high users' speed and the higher operating frequency bands (e.g. sub-$6$GHz and above-$25$GHz millimeter waves) in 5G, it is an open question whether existing 5G mobility management designs can support seamless high speed mobility. %, such that an uninterrupted very high data rate can be continued during the migration from one cell to another in the handover operation. 
Moreover, since 5G will connect a massive number of UEs/IoT devices, the network will become dense, and consequently a mobility management scheme maintaining the handover requirements for all UEs is crucial, which constitutes an open research challenge in 5G technology. %Therefore, 5G networks demand an efficient mobility management mechanism to control and manage handover operations for highly mobile devices. %Specifically, 5G and beyond network should identify the exact target cell for each connected mobile device with significantly low latency, supporting very high throughput in the network.

\subsection{Related Works}

Several existing works consider mobility in 5G networks. In~\cite{ge2016user}, the performance of user mobility in 5G small cell networks is evaluated by clustering the mobility pattern of users. The surveys in~\cite{shayea2020key,gures2020comprehensive,aljeri2020mobility} concentrate on key factors that can significantly help increase mobility management issues in 5G, where it is mentioned that smart handover approaches are required for future mobility management. To address challenges of 5G-enabled vehicular networks, the features of existing mobility management protocols are reviewed in~\cite{aljeri2020mobility}. Concentrating on a high speed train application scenario, the specifications of 5G NR are introduced in~\cite{noh2020high}, where NR design elements are also discussed. The work in~\cite{navarro2020survey} analyzes the requirements and characteristics for 5G communications, primarily considering traffic volume and network deployments.

For mobility management in 5G, a handover mechanism, that does not consider the adjustment of the TTT and hysteresis, is proposed in~\cite{choi2019generalized}. The work in~\cite{ko2017sdn} handles the mobility in 5G using a centralized software-defined networking (SDN) controller to implement the handover and location management functions. However, the centralized control can impose a communication delay during handover. While focusing only on device-to-device (D2D) communications, in~\cite{yilmaz2014smart}, several mobility management techniques are proposed, and their technical issues and expected gains are reviewed. Authors in~\cite{alsaeedy2019mobility} consider the power consumption and signaling overhead in 5G IoT devices, and accordingly handle mobility management by improving the tracking area update (TAU) approach and paging procedures. The 5G mobility management protocol proposed in~\cite{fafolahan2019seamless} is based on dividing the service area into several sub-areas, which facilitates handover in dense small cells. However, the latency between the trigger and decision in a handover is not highlighted. Based on the application-specific strategy, the work in~\cite{mumtaz2020dual} proposes a mobility management scheme supporting a data split approach between 4th Generation (4G) and 5G radio access technologies. In~\cite{kominami2017control}, a control node is introduced for managing and monitoring the autonomous distributed control in a 5G network, where the proposed control method increases the system stabilization and reduces the control plane's overhead. 
Wang \textit{et al.}~\cite{wang2017localized} present a localized mobility management (LMM) with a centralized and distributed control scheme, and show that the LMM with a centralized control mechanism has a lower handover latency and signaling cost than the LMM with distributed control.

Considering the received power from cells, in~\cite{khan2020approach}, it is shown that the selection of the base station based on the maximum received power outperforms the cell association based on the maximum signal-to-interference-plus-noise ratio (SINR). By estimating the vehicular 5G mobility across radio cells, in~\cite{labriji2021mobility}, the computing services running on mobile edge nodes are migrated for service continuity at vehicles, which effectively controls the trade-off between energy consumption and seamless computation while migrating the computing services. To save energy by turning off unused stations, a green handover procedure, that minimizes the energy consumption in 5G networks using the concept of Self-Organizing Networks (SONs), is proposed in~\cite{boujelben2017handover}.
Towards the softwarization of cellular networks, the authors in~\cite{morales2015taming} discuss 5G mobility management considering the Functionality as a Service (FaaS) platform, where maintaing a low handover latency can be a challenge. Considering a gateway selection approach, a 5G mobility management scheme based on network slicing, that supports low latency services in the closest network edge, is proposed in~\cite{heinonen2016mobility}.  
To address security issues in handovers, a distributed mobility management (DMM)-based protocol, that supports privacy, defends against redirection attacks, and provides security properties, such as mutual authentication, key exchange, confidentiality, and integrity, is proposed in~\cite{kim2020dmm}.

In the direction of dynamic handover management in 5G networks, a handover control parameter optimization mechanism for each UE, which applies a threshold-based adjustment of the TTT, is discussed in~\cite{shayea2020individualistic}; however, the control of the hysteresis is not clearly highlighted. Considering a centralized reinforcement learning agent, Yajnanarayana \textit{et al.}~\cite{yajnanarayana20205g} propose a handover mechanism for 5G networks based on measurement reports from UEs. However, due to the centralized control, the communication overhead affects the handover time. To minimize the handover failure rate, the work~\cite{mishra2020novel} initiates the handover in advance before UEs face radio link failure. Authors in~\cite{li2020beyond} consider reliable extreme mobility management in 5G and beyond, and apply the delay-Doppler domain for designing movement-based mobility management. Although the proposed scheme reduces handover failures compared to low mobility and static scenarios, the lack of dynamic adjustment of the TTT and hysteresis can degrade the handover performance in a 5G network where the signal strength varies rapidly. 

Considering challenges in handover management in 5G, an optimal gNB selection mechanism, which is based on spatio-temporal estimation techniques, is proposed in~\cite{bilen2016optimal} for intra-macrocell handovers. Due to the short wavelength, mmWave connections are easily broken by obstacles, and to address this challenge, a handover protocol is proposed in~\cite{wang2019mmhandover} for 5G mmWave vehicular networks. To address the issue of the inter-beam unsuccess handover rate in 5G networks, the proposed mechanism in~\cite{ren2019robust} designs an optimized dynamic inter-beam handover scheme. In railway wireless systems, the minimization of service interruptions during handovers is a great challenge, and accordingly a network architecture is designed in~\cite{yan2017novel} for heterogeneous railway wireless systems to achieve fast handovers in 5G. Since the Received Signal Strength Indicator (RSSI) plays a key role in fast handovers, the work~\cite{he2017adaptive} predicts the RSSI to accurately and timely trigger handovers when a mobile node is moving. In the direction of programmatically efficient management of fast handover, the network performance and monitoring can be improved further by using the SDN technology. Thus, the authors in~\cite{erel2019road} propose a SDN-based handover mechanism that triggers handovers with the help of network-centric monitoring, where an optimization approach and the shortest path considering traffic intensities of switches are used. To design a fuzzy logic based multi-attributed handover approach for 5G networks, Mengyuan \textit{et al.}~\cite{liu2018multiple} propose an optimal weight selection mechanism that considers types of services, network features, and user preferences.

Therefore, existing works do not deal with the dynamic adjustment of the TTT and hysteresis during the handover execution in 5G mobility management; however, these parameters significantly impact the successful handover execution. Specifically, based on the present network condition, the exact target cell should be identified with the appropriate adaptation of the TTT and hysteresis, such that a low handover latency and very high throughput can be achieved. %, which will be suitable for the high mobility scenarios in 5G networks. 

%An intelligent adaptation of the TTT and hysteresis can lead to a significantly low handover latency and very high throughput, which will be suitable for the high mobility scenarios in 5G networks.

%Specifically, based on the present network condition, the 5G and beyond network should identify the exact target cell for each connected mobile device with the appropriate adaptation of TTT and hysteresis, such that significantly low handover latency and very high throughput can be achieved, which will be suitable for the high mobility scenarios in 5G networks.

\subsection{Problem Statement}

%This work address the question: {\em ``How can we intelligently handle the high mobility in 5G and beyond with an adaptive selection of TTT and hysteresis?''}. 
{\em We address the problem of intelligently handling the high mobility in 5G and beyond with an adaptive selection of the TTT and hysteresis.}
To find a solution, this work targets the design of an online learning-based handover mechanism with a dynamic adjustment of the TTT and hysteresis, leading to intelligent mobility management in high speed cellular networks. The solution should be able to cope with dramatic wireless dynamics due to high mobility. Therefore, considering the specifications of 5G, the proposed model needs to follow an intelligent approach to take smart handover decisions such that delays, errors, and failures are significantly reduced in mobility management. %Moreover, due to the online learning-based automatic approach, the solution will support all usage scenarios of 5G. 
In addition, the model should be compatible with existing cellular networks. % in low mobility and will provide flexible policy for the cellular operators.  

\subsection{Proposed Approach}

%This work address the question: {\em ``How can we intelligently handle the high mobility in 5G and beyond with an adaptive selection of the TTT and hysteresis?''}. The problem statement is as follows.
 
We design a novel online learning-based approach, known as {\em \textbf{L}earning-based \textbf{I}ntelligent \textbf{M}obility \textbf{M}anagement (LIM2)}, for mobility management handling in 5G and beyond, with a dynamic adaptation of the TTT and hysteresis. LIM2 is broadly a two step approach, where in the first step, a {\em Kalman filter} is used to estimate future (a posteriori) Reference Signal Received Power (RSRP) values of the serving and neighbor cells, where the estimation is based on the measurement reports received from neighbor cells. 
Considering the predicted RSRP, in the second step, {\em state-action-reward-state-action (SARSA)} reinforcement learning is used to dynamically select the target cell for the handover. In order to maximize the cumulative reward received from an environment, SARSA decides the next action depending on the current action and the policy being used. %Therefore we apply SARSA to take the action of choosing the target cell, such that the cumulative network performance will be improved after the handover. 
During the handover, since the performance of a mobile device highly depends on the appropriate selection of the target cell among available neighbor cells, we need an on-policy based learning that can adapt the cell selection considering the present network condition, such that the cumulative network performance is improved after the handover, and therefore SARSA is an effective choice for this purpose.

Moreover, in the second step of LIM2, the {\em $\epsilon$-greedy} policy is applied as a reinforcement learning approach to dynamically choose the TTT and hysteresis based on the RSRP predicted in the first step. The $\epsilon$-greedy mechanism is an online learning approach that {\em explores} available values of a configuration and {\em exploits} the best value of a configuration considering the situation of the present execution. Since we need to explore possible available TTT and hysteresis values, and exploit the best suited values of these parameters depending on the present handover condition, we apply the $\epsilon$-greedy policy to learn about the adaptation of the TTT and hysteresis without any prior knowledge about the environment. 

\subsection{Contribution of This Work}

This work has four primary contributions as detailed below.
\begin{enumerate}
 \item We design a Kalman filter that computes the a posteriori of the RSRP values of the serving cell and possible neighbor cells for the handover. Therefore, based on the prediction, the neighbor cell that will have the highest signal quality in the future is identified as the target cell. In the high mobility scenario, the estimation of future signal quality helps take a decision on the handover in advance, such that the selected target cell is able to maintain the required network performance in high mobility scenarios.
 \item We use an online learning approach (SARSA) to dynamically select the target cell from available possible neighbor cells. For this selection, the RSRP estimated by the Kalman filter is considered, and consequently the intelligent selection of the target cell is influenced by the predicted future signal quality of neighbor cells. 
 \item We design an online learning-based mechanism for the selection of the TTT and hysteresis, considering the RSRP estimated by the Kalman filter. Based on the signal quality, these parameters are adaptively controlled in the handover execution, which results in intelligent mobility management. 
 \item We create a prototype of LIM2 by implementing it in network simulator (NS) version NS-3.33~\cite{ns3}, where the 5G NR module is plugged into \texttt{ns-3-dev}. We thoroughly evaluate the performance of LIM2 with a focus on the throughput, handover latency, and handover failure.
\end{enumerate}

\subsection{Organization of This Paper}

The rest of this paper is organized as follows. Section~\ref{sec:handover} presents an overview of 5G mobility management, along with the implications for 5G. Section~\ref{sec:model} discusses details of different modules of the proposed model. The implementation details and the performance analysis of the proposed model are presented in Section~\ref{sec:impl}. We conclude this paper in Section~\ref{sec:concl}.

\section{Overview of 5G}
\label{sec:handover}

%In this section, we discuss general specifications and implications of 5G.
In this section, we discuss the general mobility management of 5G and its implications.
%The primary advantage of 5G is that it has greater bandwidth that leads to higher download speeds, eventually up to $10$ Gbps~\cite{3GPP2,3GPP3}. 

%\subsection{Implications for 5G}

%The 5G standard offers several new features that are not supported by 4G LTE, such as dense small cells, renovated physical layer design, advanced signaling protocols, and new radio bands in the sub-6GHz or beyond-20GHz~\cite{3GPP2,3GPP3}. Since 2019, the 5G standard has been under active testing and deployment~\cite{li2020beyond}. Moreover, as reported in~\cite{li2020beyond}, it is noted that a reliable extreme mobility management in 5G will be a significant challenge because -- (1) 5G handovers follow the same design approach as 4G~\cite{3GPP1,3GPP2}, (2) 5G requires more frequent handovers due to the consideration of dense small cells that can use high carrier frequencies~\cite{gures2020comprehensive,navarro2020survey}, (3) frequent handovers increase the rate of handover failures, and (4) although 5G improves the reliability by refining its physical layer (e.g. more reference signals and polar coding)~\cite{3GPP3}, the standard is still based on orthogonal frequency-division multiplexing (OFDM), and thus has several issues such as high peak-to-average power ratio (PAPR), time and frequency synchronization, and high sensitivity to the inter-modulation distortion. %and suffers from the issues related to OFDM.

\subsection{5G Mobility Management}

High speed mobility management is governed by an efficient handover operation which is primarily based on the signal quality of a cell.

%\noindent{\textbf{Signal power and channel quality:}} 
\subsubsection{Signal power and channel quality}
When a handover is initiated, a measurement report is sent from a mobile device to the serving cell base station (or gNB)~\cite{3GPP1,3GPP2}. The measurement report consists of several channel quality related parameters, such as the RSRP, the Reference Signal Received Quality (RSRQ), the RSSI, etc. These parameters are measured both for the serving cell and neighboring cells. The RSRP represents the linear average power of the reference signal, which is measured over the full bandwidth expressed in resource elements (REs). The RSRP is the most significant measurement considered for handover. UEs usually measure the RSRP based on the Radio Resource Control (RRC) message from the base station. On the other hand, the RSRQ considers the RSRP with RSSI and the number of resource blocks. The RSRQ determines the quality of the reference signal received from a base station. The RSRQ measurement supplies additional information when a reliable handover or cell reselection decision cannot be made based on the RSRP.

\begin{figure}[!t]
 \centering
 \includegraphics[width=\linewidth]{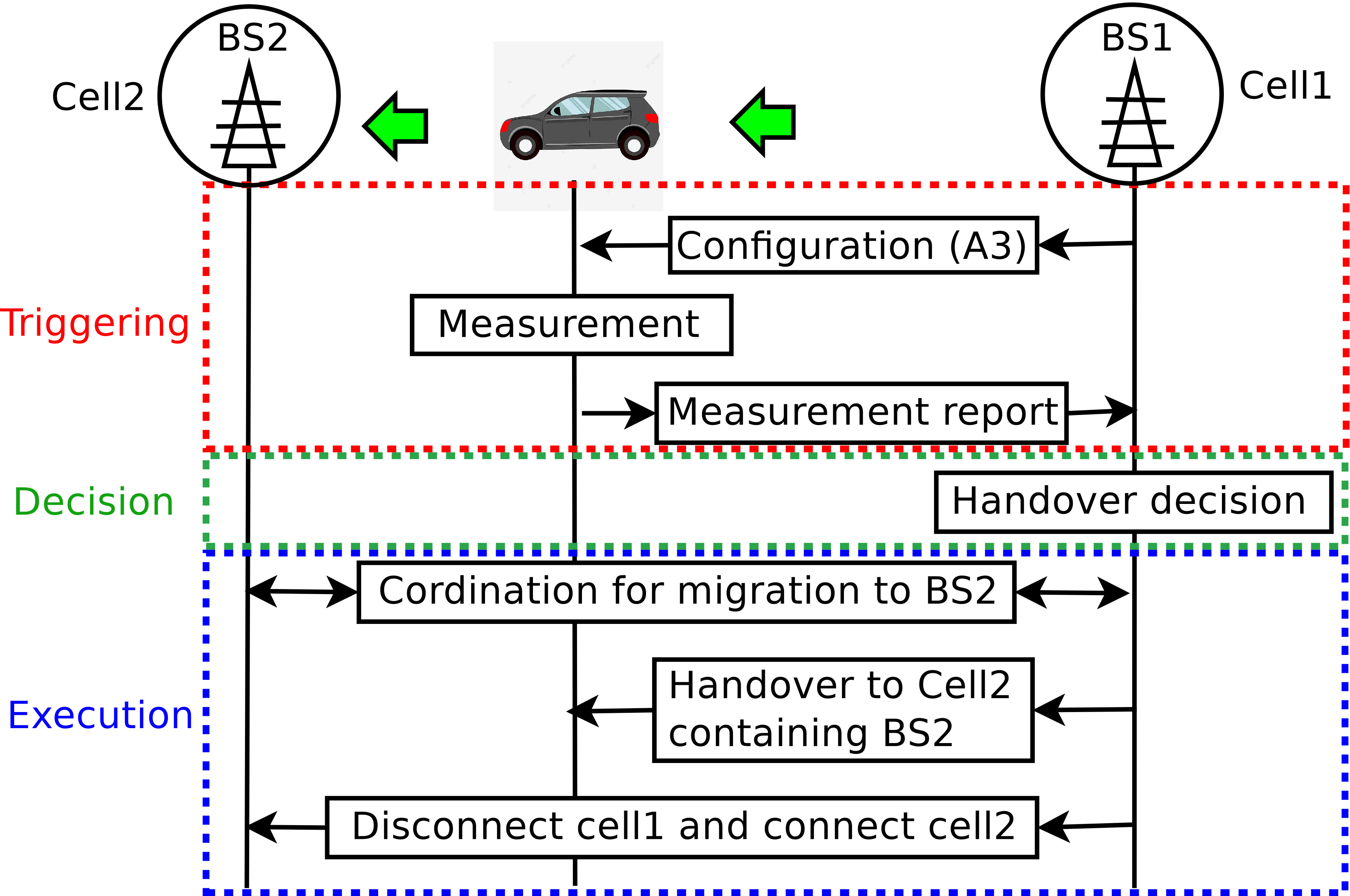}
 \caption{Procedural view of handover}
 \label{fig:ho-proc}
\end{figure}

%\noindent{\textbf{Handover operation:}} 
\subsubsection{Handover operation}
In mobile networks, a base station acts as a fixed transceiver and is the primary communication point for wireless mobile client devices. %To enable ubiquitous access, base stations are deployed in 4G/5G to cover different areas. 
Basically, a base station manages the communication in a cell and sometimes multiple base stations may run under different frequency bands and different coverage with the help of separate antennas. As a mobile device enters into a new cell's coverage, after leaving the old cell, the mobile device will be migrated to the new cell to retain its network access. Consequently, the control of the communication will be transferred from one cell to another using the {\em handover} or {\em handoff} mechanism.

As illustrated in Fig.~\ref{fig:ho-proc}, the handover operation has three phases -- {\em triggering}, {\em decision}, and {\em execution}~\cite{3GPP1,3GPP2}. The handover starts with the triggering phase with the serving cell asking a mobile device for the measurement report to measure the signal strength of neighbor cells, where standard triggering criteria are shown in Table~\ref{table:ho-trigger}. After receiving the mobile device's feedback, the decision phase is started and the serving cell takes the handover decision and identifies the target cell for the migration based on the triggering criteria. For this purpose, the serving cell may reconfigure the mobile device for more feedback. After the completion of a handover decision, the execution phase, where the target cell is coordinated and the handover command is transmitted to the mobile device, begins. Then, the mobile device is disconnected from the serving cell and connected to the target cell.

\begin{table}
\caption{Handover triggering criteria~\cite{3GPP1,3GPP2}}
\centering
\begin{tabular}{|p{1cm}|p{2cm}|p{4.5cm}|}
\hline
\textbf{Event} & \textbf{Criteria} & \textbf{Explanation} \\
\hline
%\hline
A1 & $R_s > \Delta_{A1}$ & Serving cell's RSRP is better than a threshold \\
\hline
A2 & $R_s < \Delta_{A2}$ & Serving cell's RSRP is worse than a threshold \\
\hline
A3 (A6) & $R_n > R_s + \Delta_{A3}$ & Neighbor cell is better than serving cell with a offset \\
\hline
A4 (B1) & $R_n > \Delta_{A4}$ & Neighbor cell is better than a threshold \\
\hline
A5 (B2) & $R_s < \Delta_{A5}^1$, $R_n > \Delta_{A5}^2$ & Serving cell is worse than a threshold value, and neighbor cell is better than a threshold\\
\hline
\end{tabular} 
\label{table:ho-trigger}
\end{table}

\begin{figure}[!t]
 \centering
 \includegraphics[width=0.8\linewidth]{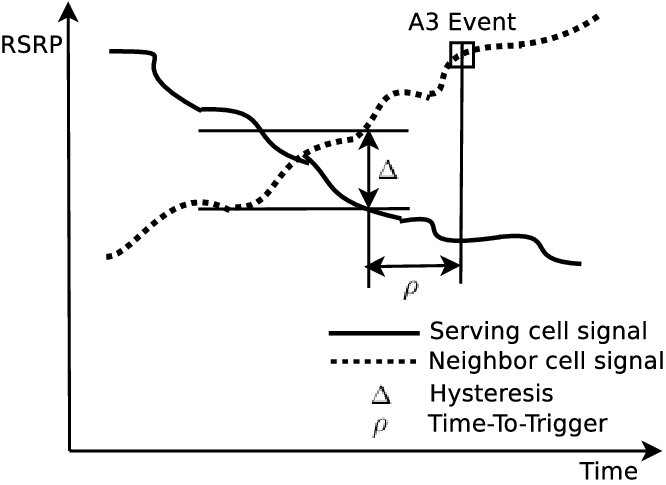}
 \caption{Illustration of handover mechanism}
 \label{fig:ho}
\end{figure}

%\noindent{\textbf{Time-to-Trigger and hysteresis:}} 
\subsubsection{Time-to-Trigger and hysteresis}
Fig.~\ref{fig:ho} illustrates the Time-to-Trigger (TTT) and hysteresis margin, which are important parameters supported by the Long-Term Evolution (LTE) and 5G standards to trigger the handover procedure and choose the target cell~\cite{3GPP1,3GPP2}. When the triggering criteria is fulfilled for a TTT interval, the handover mechanism is initiated. The TTT decreases unnecessary handovers, and thus effectively avoids ping-pong effects due to the repeated movement of mobile devices between a pair of cells (serving and target cells). In Fig.~\ref{fig:ho}, it is noted that the A$3$ handover event is initiated after the TTT interval and should maintain a hysteresis margin based on the RSRP/RSRQ values of the serving and target cells.

\subsection{Implications for 5G}

The 5G standard offers several new features that are not supported by 4G LTE, such as dense small cells, renovated physical layer design, advanced signaling protocols, and new radio bands in the sub-6GHz or beyond-20GHz~\cite{3GPP2,3GPP3}. Since 2019, the 5G standard has been under active testing and deployment~\cite{li2020beyond}. Moreover, as reported in~\cite{li2020beyond}, it is noted that a reliable extreme mobility management in 5G will be a significant challenge because -- (1) 5G handovers follow the same design approach as 4G~\cite{3GPP1,3GPP2}, (2) 5G requires more frequent handovers due to the consideration of dense small cells that can use high carrier frequencies~\cite{gures2020comprehensive,navarro2020survey}, (3) frequent handovers increase the rate of handover failures, and
(4) although 5G improves the reliability by refining its physical layer (e.g. more reference signals and polar coding)~\cite{3GPP3}, the standard is still based on orthogonal frequency-division multiplexing (OFDM), and thus has several issues such as high peak-to-average power ratio (PAPR), time and frequency synchronization, and high sensitivity to the inter-modulation distortion. %and suffers from the issues related to OFDM.

\section{LIM2: Model Formulation}
\label{sec:model}

\begin{figure}[!t]
 \centering
 \includegraphics[width=\linewidth]{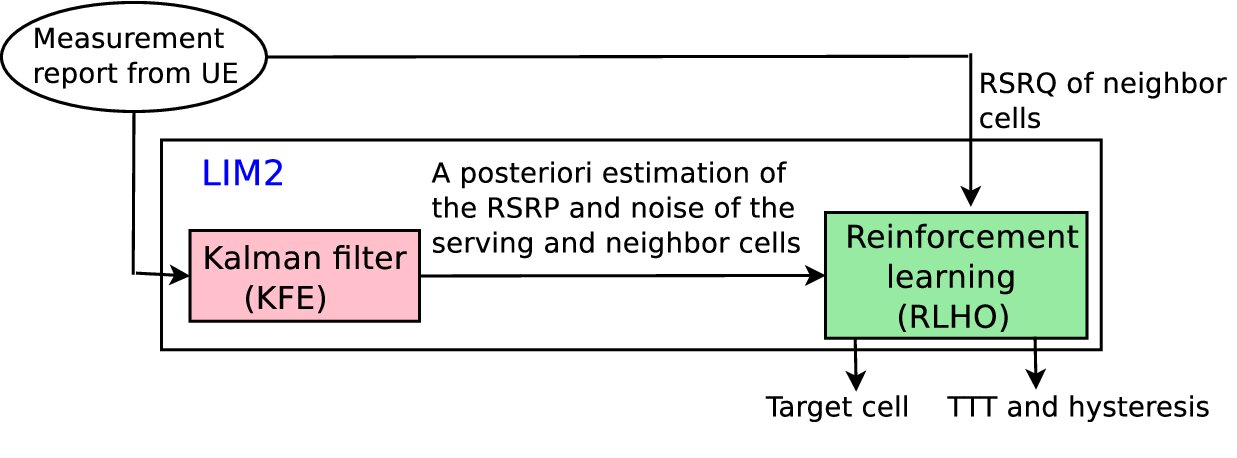}
 \caption{LIM2: System model overview}
 \label{fig:overall}
\end{figure}

In this section, we present the proposed model. To this end, Fig.~\ref{fig:overall} shows the overall system model of LIM2, which consists of two modules -- {\em (i) the Kalman filter based RSRP estimation (KFE)} and {\em (ii) the reinforcement learning-based handover (RLHO)}. The measurement report is the input to the model, and based on the report, the KFE module estimates the a posteriori of the RSRP and noise of the serving and neighbor cells. The RLHO module runs SARSA by considering the output of the KFE module and the RSRQ of neighbor cells, and then selects the target cell for a handover. In addition, the RLHO applies the $\epsilon$-greedy policy to adaptively choose the TTT and hysteresis value for the handover. Details of each of the modules (KFE and RLHO) are discussed next.

\subsection{Kalman Filter based RSRP Estimation}

The Kalman filter~\cite{kalman1960new} follows a recursive estimation, i.e., to compute the next state, only the previous state and the present measurement are required.
The reasons for using the Kalman filter are:
\begin{itemize}
 \item As a recursive model, the Kalman filter does not need to know the entire history of measurements. The Kalman filter only requires information about the last measurement to estimate the desired output~\cite{kalman1960new}. As a result, the response time is reduced and the memory space is saved.
 \item The estimation of unknown variables using the Kalman filter tends to be more accurate than that based on a single measurement~\cite{olfati2009kalman}.
 \item The Kalman Filter optimizes the estimation error; specifically, the mean squared error is minimized by it while considering systems with Gaussian noise~\cite{pishdad2020analytic}.
 \item The Kalman filter has a promising ability to estimate (track) the system states (parameters) from noisy measurements~\cite{diversi2005kalman}. Therefore, it can be used to estimate the signal strength of the channel, and accordingly track the variation of channel quality, considering the noise factor.
\end{itemize}
On the other hand, classical machine learning techniques need historical data for learning purposes, where the estimation is based on the quality and size of the data volume. However, the signal strength of channels varies quickly, and thus determining the RSRP based on a long series of past measurements will not be fast enough to cope with a wireless environment. In addition, for the estimation of RSRP values, appropriate and sufficient historical data is required such that RSRP values can be dynamically computed with minimum errors in different network conditions. Moreover, due to the requirement of a large volume of historical data, classical machine learning schemes require sufficient memory space, and it is also challenging to add appropriate noise as latent factors in the past measurements.

Let $Y_{RSRP,k}$ and $Y_{RSRP,k-1}$ be the RSRP values measured at time $k$ and $(k-1)$, respectively. Let $N_e$ be the environment noise, which may come from other radio frequency (RF) devices, such as WiFi, power generator, motor, microwave oven, etc. The mobile network is a time-varying system, where the environment and measurement noises are important factors. These two parameters affect the effective signal strength of a cell. Channel quality estimation is crucial to take the appropriate decision in a handover. Therefore, it is required to improve the channel quality estimation by getting rid of the measurement noise. Since the Kalman filter has a promising ability to estimate (track) the system states (parameters) from noisy measurements, we can design a Kalman filter to estimate the signal strength of the channel, and accordingly track the variation of channel quality, considering the noise factor. The Kalman filter may give a higher accuracy with fast tracking ability. In our model, the Kalman filter addresses the variation of RSRP ($Y_k$) and environment noise ($N_e$) by filtering the value of the measurement noise. Therefore, the system can be modeled as
\begin{equation}\label{eq:k1}
\begin{aligned}
 Y_{RSRP,k} &= Y_{RSRP,k-1} + w_{s,k} \\
 N_{e,k} &= N_{e,k-1} + w_{e,k}.
\end{aligned}
\end{equation}
Here, $w_{s,k}$ and $w_{e,k}$ denote the impact of path loss, fading, and shadowing on the RSRP, and the environment noise, respectively. $w_{s,k}$ and $w_{e,k}$ are assumed to be independent and follow Gaussian distributions with zero mean. 
The RSRP and environment noise are correlated since the effective value of the RSRP decreases as the environment noise increases and vice versa. Thus, the likelihood of the measured RSRP depends on the environment noise. At time $k$, the correlation between the RSRP and environment noise is captured by a {\em covariance matrix} denoted by $\mathbf{P}_k$. 
Eqn.~(\ref{eq:k1}) can be represented in a vector form, which leads to a state evolution equation as
\begin{equation}\label{eq:k2}
\mathbf{x}_k = \mathbf{F}_k \cdot \mathbf{x}_{k-1} + \mathbf{w}_k.
\end{equation}
Here, $
\mathbf{x}_k=\begin{bmatrix}
Y_{RSRP,k}\\
N_{e,k}
\end{bmatrix}$ denotes the estimated state value of the dynamic system. We need to predict the next state (at time $k$) based on the present state (at time $k-1$). $\mathbf{F}_k=\begin{bmatrix}1 & 0\\ 0 & 1\end{bmatrix}$ is the prediction matrix and $\mathbf{w}_k=\begin{bmatrix}w_{s,k}\\ w_{e,k}\end{bmatrix}$. The prediction matrix is used to estimate the next state. In wireless signal models, the state variable can be measured by the radio-frequency integrated circuit (RFIC) and the state variable is associated with the measurement noise. Let $\mathbf{Q}_k$ be the covariance of the zero mean Gaussian distribution followed by $\mathbf{w}_k$, and thus $\mathbf{w}_k \sim \mathcal{N}(0,\mathbf{Q}_k)$.

In the RSSI and signal-to-noise ratio (SNR)-based model, it can be seen that the state variable can be observed by the RFIC directly and is subject to measurement noise, in which the internal noise is the main contributor. The internal noise is defined as the noise that is added to the signal after it is received. Thus, the observation equation is represented as
\begin{equation}\label{eq:k3}
\mathbf{z}_k = \mathbf{H}_k \cdot \mathbf{x}_k + \mathbf{v}_k.
\end{equation}
Here, $\mathbf{z}_k$ is the observed or measured value of the state of the system. %and $z_k$ maps the estimated state into the observed state. 
$\mathbf{H}_k=\begin{bmatrix}1 & 0\\ 0 & 1\end{bmatrix}$ is the measurement matrix and $\mathbf{v}_k$ denotes the measurement noise or observed noise, which is assumed to follow a zero mean Gaussian distribution with covariance $\mathbf{R}_k$, i.e. $\mathbf{v}_k \sim \mathcal{N}(0,\mathbf{R}_k)$.

The Kalman filter is conceptualized in two phases -- {\em ``Predict''} and {\em ``Update''}~\cite{kalman1960new}. In the predict phase, the previous state estimate is used to predict the state for the current timestep. Although the predicted state is an estimation of the state at the present timestep, it does not consider information observed from the current timestep. Thus, the predicted state is also called the a {\em priori} state. In the update stage, the current observation information is combined with the present {\em a priori} prediction to refine the state estimate. This improved state estimate is known as the {\em a posteriori} estimate of the state.  

\begin{figure}[!t]
 \centering
 \includegraphics[width=0.8\linewidth]{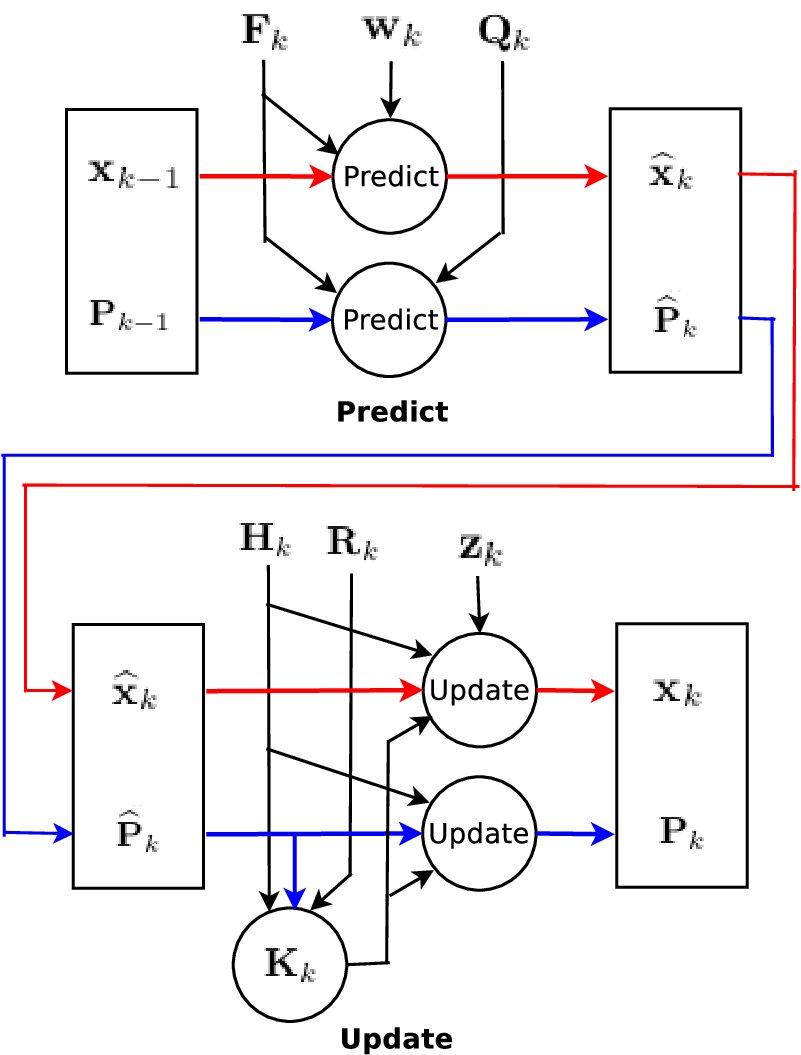}
 \caption{Information flow in the Kalman filter in KFE module}
 \label{fig:KFE}
\end{figure}

Let $\widehat{\mathbf{x}}_k$ and $\mathbf{x}_k$ be the {\em a priori} and {\em a posteriori} estimation of the state, respectively. Let $\widehat{\mathbf{P}}_k$ represent the {\em a priori} estimate error covariance matrix and $\mathbf{P}_k$ be the {\em a posteriori} estimate error covariance matrix. Assume that $\mathbf{K}$ denotes the Kalman gain which is the relative weight assigned to the current state estimate and measurements. $\mathbf{K}$ can be tuned to obtain a particular performance. When the Kalman gain is high, more weight is given to the recent measurements, and consequently these measurements are followed by the filter more reponsively. %However, when the gain is low, the model predictions are performed more closely. 
As a result, a high gain results in a frequent jump in the estimation, whereas a low gain smooths out the noise but decreases the responsiveness. At any time instant $k$, the associated two distinct phases of the Kalman filter are defined in what follows~\cite{kalman1960new}. 

\noindent{\textbf{Predict:}} In this phase, the following estimations are performed:
\begin{itemize}
 \item \textit{Prior state estimate:} $\widehat{\mathbf{x}}_k=\mathbf{F}_k \cdot \mathbf{x}_{k-1}+\mathbf{w}_k$
 \item \textit{Prior error covariance estimate:} $\widehat{\mathbf{P}}_k=\mathbf{F}_k \cdot \mathbf{P}_{k-1} \cdot \mathbf{F}^\intercal_k+\mathbf{Q}_k$
\end{itemize}

\noindent{\textbf{Update:}} In this phase, the following steps are performed:
\begin{itemize}
 \item \textit{Kalman gain:} $\mathbf{K}_k=\widehat{\mathbf{P}}_k \cdot \mathbf{H}^\intercal_k \cdot \mathbf{H}_k \cdot \widehat{\mathbf{P}}_k \cdot \mathbf{H}^\intercal_k+\mathbf{R}^{-1}_k$
 \item \textit{Posterior state update:} $\mathbf{x}_k=\widehat{\mathbf{x}}_k+\mathbf{K}_k \cdot \mathbf{z}_k - \mathbf{H}_k \cdot \widehat{\mathbf{x}}_k$
 \item \textit{Posterior error covariance update:} $\mathbf{P}_k=(\mathbf{I}-\mathbf{K}_k \cdot \mathbf{H}_k) \cdot \widehat{\mathbf{P}}_k$
\end{itemize}
Fig.~\ref{fig:KFE} shows the predict and update mechanisms in the KFE module.
Since the converged value of $\mathbf{K}$ is not affected by the initial value of $\mathbf{P}$, we can use a non-zero matrix as the initial value of $\mathbf{P}$, and $\mathbf{K}$ automatically converges to the final value. 

The state value $\mathbf{x}_k$ is used in the next module, where the handover is performed based on reinforcement learning. This learning technique and details of the handover mechanism are discussed next.

\subsection{SARSA-Based Reinforcement Learning}

A reinforcement learning (RL)~\cite{andrew1999reinforcement,zhang2020learning} model is based on the following parameters:
\begin{itemize}
 \item \textit{Set of states:} A state is used to describe the current situation of a system for a given environmental condition. Let the set of states be $S$.
 \item \textit{Set of actions:} The new state (next state) is obtained by applying an action on the present state. Let $A$ denote the set of available actions.
 \item The probability of state transition from state $s_k$ (at time $k$) to state $s_{k+1}$ under action $a_k$. Let $Pr_{a_k}(s_k,s_{k+1})$ denote the probability.
 \item \textit{Policy:} A policy defines a set of rules that are followed by the RL agent to determine the action for the current state.
 \item \textit{Reward:} A reward, generally defined by a scalar quantity, is a return value given by the environment for changing the state of the system.
\end{itemize}
\begin{figure}[!t]
 \centering
 \includegraphics[width=\linewidth]{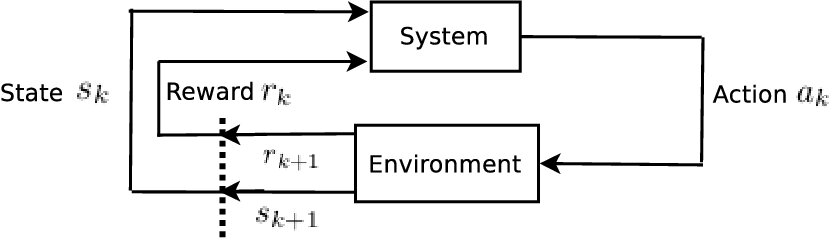}
 \caption{Reinforcement learning framework}
 \label{fig:rl}
\end{figure}
Fig.~\ref{fig:rl} shows a basic RL model that has two modules -- (i) system and (ii) environment. At any time instance, the system belongs to a state. Based on a policy, the system applies an action and changes the state. Consequently, the environment returns a reward for the change of the state.

SARSA~\cite{aslam2019optimal} is an on-line RL policy, where an agent learns the policy value and the associated action in each state transition. %by using the concept of {\em exploration} and {\em exploitation}. In exploration, an action is chosen randomly from a set of available actions; whereas, in exploitation, the action, which has provided the best reward in the past, is chosen by considering the present context. 
A SARSA agent interacts with an unexplored environment in discrete time instances and gets knowledge about the environment, such that the cumulative reward is maximized. %SARSA is known as an online-learning approach since the policy is updated based on the action taken at each step. 

\subsubsection{Components of SARSA}
\label{sec:sarsa}

At time $k$, let $s_k$ be the current state and $a_k$ be the action that changes the state $s_k$ to $s_{k+1}$. Let $r_k$ and $r_{k+1}$ be the rewards associated with $s_k$ and $s_{k+1}$, respectively. Assume that $\Pi$ denotes the policy that determines the action that needs to be applied on $s_k$ to reach the next state $s_{k+1}$, i.e., $\Pi(s_k)=a_k$. Thus, the policy $\Pi$ helps the system move to the new state $s_{k+1}$ and obtain the associated reward $r_{k+1}$. This state transition is represented as $(s_k,a_k,r_{k+1},s_{k+1})$. Now, in state $s_{k+1}$, the learning agent again uses the policy $\Pi$ and finds the action suitable for $s_{k+1}$. Let $a_{k+1}$ denote the action in state $s_{k+1}$, and thus in this state, the policy can be defined as $\Pi(s_{k+1})=a_{k+1}$.
These two consecutive transitions are generally represented by the quintuple $(s_k,a_k,r_{k+1},s_{k+1},a_{k+1})$. This quintuple signifies -- state-action-reward-state-action, i.e., ``SARSA''.

\noindent{\textbf{Action-value-function:}} In a state, the action-value-function is used to compute the expected utility for an action taken by the agent. Specifically, this function is a quantitative measure of the state-action combination. For state $s$ and action $a$, let $Q(s,a)$ be the action-value function which is also known as {\em Q-function} or {\em Q-value}. $Q(s,a)$ is defined as
\begin{equation}\label{eq:action-value}
Q(s,a)=\sum_{i=1}^{d}\theta_i\phi_i(s,a).
\end{equation}
Here, $\theta$ is a weight factor with $0\le\theta\le 1$. Let $S$, $A$, and $R$ denote the set of states, actions, and rewards, respectively. Therefore, in general, the Q-value can be represented as
$$ Q: S \times A \mapsto \mathbb{R}. $$ 

At the beginning, the Q-value is selected by the designer and can return an arbitrary value. After applying an action in a state (present state), the system transits to a new state (next state) and the Q-value is calculated for the present state. The Q-value gets updated for a state when it is considered as the present state. Therefore, the value iteration, which makes an update on the old Q-value by considering the current information related to the taken action, is the core component in SARSA. The update of the Q-value is defined as
\begin{equation}\label{sarsa}
Q(s_k,a_k) \leftarrow Q(s_k,a_k)+\alpha[r_{k+1}+\gamma Q(s_{k+1},a_{k+1})-Q(s_k,a_k)].
\end{equation}
In Eqn.~(\ref{sarsa}), the parameters $\alpha$ and $\gamma$ are known as learning rate and discount factor, respectively. These are defined in what follows.
\begin{itemize}
 \item \textit{Learning rate ($\alpha$):} This factor determines to what extent the newly acquired information will override the old information. When $\alpha=0$, the agent does not learn anything from the environment; whereas, the agent considers the most recent information only when $\alpha$ is set to $1$. In practice, a constant value, such as $0.1$, is used as the learning rate since the learner can be assigned a significant time to obtain information about the environment.
 \item \textit{Discount factor($\gamma$):} This factor finds the significance of future rewards. When $\gamma=0$, the agent considers the current reward only. The agent strives for a long-term high reward value as $\gamma$ reaches $1$. If $\gamma$ meets or exceeds $1$, the Q-value may diverge. %To impose a balance between the current and future reward value, $\gamma$ is set to $0.5$.
\end{itemize}

Next, we discuss details of the application of SARSA in our proposed model.

\subsection{Application of SARSA in LIM2}

Based on the management report sent from a UE, the serving cell takes a decision on the handover. If it is required, the serving cell initiates it by requesting the target cell for the handover. The serving cell runs the proposed SARSA model to take the handover decision. In this context, the state, action, and reward associated with our SARSA model are discussed in what follows.
\begin{itemize}
 \item \textbf{State:} The state represents a cell. At any time $k$, state $s_k$ represents the serving cell of a UE. The state $s_{k+1}$ is the neighbor cell (or the next state) considered at time $k$, which may be chosen as the target cell at time $(k+1)$ for a handover.  
 \item \textbf{Action:} At time $k$, the {\em action} $a_k$ denotes the migration from the serving cell ($s_k$) to a neighbor cell ($s_{k+1}$) that is chosen as the target cell (next state) for a handover. Therefore, the handover will be performed by shifting the control from state $s_k$ to $s_{k+1}$. Hence, $a_k$ updates the Q-value of $s_k$.
 \item \textbf{Reward:} %The target cell should be chosen such that the RSRP will be maximized and the environment noise will be minimized. 
The reward $r_{k+1}$ of $s_{k+1}$ is represented by the RSRQ of the neighbor cell. %Since the RSRQ combines the RSRP and RSSI with the number of resource blocks, a higher RSRQ signifies a higher throughput. Therefore, we consider the RSRQ as the reward in our SARSA model, where the objective is to increase the throughput under mobility such that the high performance of 5G is maintained. The RSRQ is also useful to determine the target cell for the handover when the RSRP is not sufficient to make the handover decision. Moreover, when the RSRP values of two cells are similar, the RSRQ becomes crucial to choose the target cell. 
The RSRQ can be obtained from the measurement report sent by a UE. 

\end{itemize}

\subsubsection{Generation of Q-Values}

As we discussed in Section~\ref{sec:sarsa}, the Q-value, measured by (\ref{sarsa}), produces the expected utility for applying an action in a state. Thus, to compute the Q-value, we need to find -- the state, action, and reward. Let $x_k$ represent the scalar combination of the RSRP and environment noise, which are extracted from $\mathbf{x}_k$. Since both the RSRP and environment noise are crucial in handover, we consider the combination of the RSRP and environment noise to represent the Q-value, i.e., $\text{Q-value}=x_k$.

\noindent{\textbf{Types of Q-value:}} Two types of Q-value are defined for each state -- initial and final values, which are defined as follows.
\begin{enumerate}
 \item $Q^{init}$: The last updated Q-value of a cell during the last handover. % is considered as $Q^{init}$.
 \item $Q^{final}$: The updated Q-value of a cell, which is based on $Q^{init}$, the reward, and $x_k$ values of a cell and its neighbor cell.
\end{enumerate}
The serving cell updates the Q-value by considering its Q-value during the last handover, the $x_k$ values of the serving and target cells, and the associated reward. Therefore, to perform the handover, each cell maintains a Q-value which is updated when the measurement report is sent by a UE.

\noindent{\textbf{Update of Q-value:}}
At time $k$, let the last updated Q-value of the serving cell be $Q(s_k,a_k)^{init}$. Let the values of the RSRP and environment noise of the serving and neighbor cells be $x_k^{srv}$ and $x_k^{nbr}$, respectively. Also, let the updated Q-value of the serving cell be $Q(s_k,a_k)^{final}$ which is used to migrate to the target cell. In this context, $a_k$ specifies the action for the handover from the serving cell to one of the neighbor cells (the target cell). Assume that the Q-value of the neighbor cell is represented by $Q(s_{k+1},a_{k+1})$, and thus we have $Q(s_{k+1},a_{k+1})=x_k^{nbr}$. Therefore, $s_{k+1}$ signifies the state related to the target cell that will be reached after action $a_k$. Since we consider the RSRQ as reward, at time $k$, let $V_{RSRQ,k}^{nbr}$ be the RSRQ of a neighbor cell, and therefore $r_{k+1}=V_{RSRQ,k}^{nbr}$.
%The associated reward $r_{k+1}$ of $s_{k+1}$ is calculated based on Shannon capacity which is used to find out the theoretical upper bound of data rate in a noisy channel. Therefore, $r_{k+1}$ is represented as follows.
%The associated reward $r_{k+1}$ of $s_{k+1}$ is represented by RSRQ of the neighbor cell. RSRQ is useful to determine the target cell for the handover when RSRP is not sufficient enough to make the handover decision. Since RSRQ combines RSRP and RSSI with the number of resource blocks, a higher RSRQ value signifies more expected throughput. Moreover, when RSRP values of two cells are same, RSRQ becomes crucial to choose the target cell. RSRQ can be obtained from the measurement report sent by a UE. At time $k$, let $V_{RSRQ}^{nbr}$ be RSRQ of the neighbor cell, and therefore $r_{k+1}=V_{RSRQ}$.
%\begin{equation}\label{throughput} 
%r_{k+1} = W\log(1+S).
%\end{equation}
%Here, $W$ is channel bandwidth (in Hz) of one PRB and $S$ denotes the SNR of the channel. $S$ is represented as a linear power ratio. 
Now, following (\ref{sarsa}), we define $Q(s_k,a_k)^{final}$ as 
%$$Q(s_k,a_k) \leftarrow Q(s_k,a_k)+\alpha[r_{k+1}+\gamma Q(s_{k+1},a_{k+1})-Q(s_k,a_k)]$$
\begin{equation}\label{sarsa2}
Q(s_k,a_k)^{final} \leftarrow Q(s_k,a_k)^{init}+\alpha[V_{RSRQ,k}^{nbr}+\gamma x_k^{nbr}
-x_k^{srv}].
\end{equation}
In (\ref{sarsa2}), it is noted that both the serving and neighbor cells' $x_k$ values, i.e. $x_k^{srv}$ and $x_k^{nbr}$, respectively, are used to compute $Q(s_k,a_k)^{final}$, such that both the serving and neighbor cells' predicted signal qualities can be utilized for the handover decision, along with $Q(s_k,a_k)^{init}$ of the last handover in that serving cell and the RSRQ of the neighbor cell. Therefore, all the aforementioned crucial handover related factors are combined into a single Q-value, leading to efficient mobility management. A higher Q-value indicates that both the RSRP and RSRQ are high with a low environment noise. Therefore, the objective is to increase the Q-value.  

Since $x_k$ is a combination of the RSRP and environment noise, these two parameters are normalized between $[0,1]$ and added to quantify $x_k$ with a scalar value. Then, $x_k$ is further normalized between $[0,1]$. Similarly, the RSRQ and $r_{k+1}$ are also normalized between $[0,1]$. Therefore, the Q-value is a normalized value expressed between $[0,1]$. For the normalization, we use the \textit{sigmoid logistic function}~\cite{HanSigmoid:1995} since it is a bounded differentiable real function. The sigmoid function is defined for all real values with a positive derivative at each point. The sigmoid function $h(a)$ of variable $a$ is defined as $h(a)=\frac{1}{1+e^{-a}}$. Thus, as per the sigmoid function, the normalized value of $x_k$ is $h(x_k)=\frac{1}{1+e^{-x_k}}$.

\noindent{\textbf{Reason of using RSRQ as reward:}} The RSRQ combines the RSRP and RSSI with the number of resource blocks and is defined using the interference power. The information in a weak signal can be extracted from a high RSRQ-based connection because of its minimal noise. Therefore, a higher RSRQ can lead to a higher throughput and consequently reduced packet loss rate, block error rate, etc~\cite{raida2018deriving}. Thus, the use of a single parameter, i.e. the RSRQ, can help capture the overall performance of a handover decision instead of using multiple metrics, such as throughput, packet and block error rate, etc., since individually, these metrics are not sufficient for an efficient handover decision. Since a single parameter, that reflects the overall network performance, is required to define the reward in SARSA, we consider the RSRQ as the reward in our SARSA model, where the objective is to improve the performance of 5G under mobility. The RSRQ is also useful to determine the target cell for the handover when the RSRP is not sufficient to make the handover decision. Moreover, when the RSRP values of two cells are similar, the RSRQ becomes crucial to choose the target cell. Hence, the consideration of the RSRQ as the reward also helps integrate the RSRQ with the RSRP in order to perform the handover operation.

Next, we describe the $\epsilon$-greedy policy which is another RL-based online learning approach used in the RLHO module.

\subsection{$\epsilon$-greedy Policy}

The $\epsilon$-greedy~\cite{Epsilon-greedy} is a well known policy in reinforcement learning. The $\epsilon$-greedy policy handles the trade-off between {\em exploration} and {\em exploitation}. The $\epsilon$-greedy introduces a parameter, known as exploration probability, to impose the rate of exploration. At time $k$, we define $\epsilon_k$ as
\begin{equation}\label{epsilon}
\epsilon_k = min(1,rN/k^2).
\end{equation}
Here, $N$ is the sum of available number of the TTT and hysteresis defined by the 5G NR standard~\cite{3GPP1,3GPP2}. $r>0$ is a parameter that adjusts exploration. In the $\epsilon$-greedy policy, the exploration and exploitation are defined as follows. 
\begin{itemize}
 \item \textit{Exploration:} In exploration, an action is selected randomly, where at time $k$, the probability of exploration is $\epsilon_k$.
 \item \textit{Exploitation:} In exploitation, the action that has produced the best reward so far is selected. At time $k$, the exploitation probability is $(1-\epsilon_k)$.
\end{itemize}
The exploration and exploitation can be represented as a {\em strategy}, and therefore, at time $k$, the {\em strategy} is defined as
$$ strategy = \epsilon_k \times explore + (1-\epsilon_k)\times exploit. $$

\subsection{Q-Table for Adaptation of TTT and Hysteresis}

%We apply $\epsilon$-greedy mechanism as the policy for selecting TTT and hysteresis, and for this purpose, the learning agent of our SARSA model is used. 
Let the TTT and hysteresis values be $\rho$ and $\Delta$, respectively. Moreover, let {\em Q-Table} be a table that keeps information related to the selected $(\rho,\Delta)$ and the associated Q-value ($Q(s_k,a_k)^{final}$). Thus, the {\em Q-Table} is represented as Q-Table=$\{(\rho,\Delta),Q(s_k,a_k)^{final}\}$. In this context, $Q(s_k,a_k)^{final}$ is used to select the $(\rho,\Delta)$ pair such that $Q(s_k,a_k)^{final}$ is maximized in the exploitation phase of the $\epsilon$-greedy approach. In a high mobility scenario, $(\rho,\Delta)$ should not be large and it needs to be chosen such that, after the handover, a high throughput can be maintained in the target cell. Therefore, we use the Q-value to choose $(\rho,\Delta)$ because a higher Q-value implies a higher RSRP and RSRQ, and therefore a higher expected throughput in the target cell. 

At the beginning of the execution of LIM2, the RSRP and RSRQ values of all the neighbor cells are extracted from the measurement report along with the RSRP value of the serving cell. Based on the RSRP values, the state value $x_k$ is computed for the serving cell and all the neighbor cells using the Kalman filter. Then, based on the RSRQ and $x_k$ values, the Q-value is calculated for all the neighbor cells; and one of the neighbor cells, which provide the maximum Q-value, is identified as the target cell. Since the Q-value is computed based on the output of the Kalman filter and the RSRQ of neighbors, the selected target cell has the highest possibility to have the best future signal quality among the other neighbors after handover. After determining the target cell, the TTT and hysteresis margins are dynamically selected only for the target cell by applying $\epsilon$-greedy policy. Therefore, we can assume that in LIM2, SARSA acts as an implicit filter to intelligently choose the best possible target cell such that we need to compute the TTT and hysteresis only for the target cell.

In order to collect the measurement reports, the proposed mechanism initially determines values of the TTT and hysteresis margin by following the 5G standard. Since the TTT and hysteresis are determined based on the procedure as defined by the 5G standard, before collecting the measurement reports, the determination of the TTT and hysteresis is not an online learning-based approach in the proposed mechanism. However, after determining the target cell, the proposed mechanism further computes the TTT and hysteresis for the target cell using a SARSA-based RL approach, such that the handover delay and ping-pong effect are reduced and redundant handovers are eliminated. Therefore, the final handover decision is intelligently handled by LIM2, leading to adaptive mobility management. Since LIM2 additionally calculates the TTT and hysteresis after determining the target cell, it does not require any modifications to the 5G standard.

\subsection{Execution Details of LIM2}

The proposed LIM2 uses SARSA-based online learning as an RL learning mechanism. Here, SARSA does not require any predefined dataset, training, or testing. Thus, it is not required to train and test the proposed LIM2 model. However, to select an action, initially we need information related to the selection of $(\rho,\Delta)$ and the associated Q-value ($Q(s_k,a_k)^{final}$), and that information is stored in the Q-Table. Initially, the table is empty and as the execution time of LIM2 increases, the Q-Table is populated using the $\epsilon$-greedy policy, which has two phases -- exploration and exploitation. In the initialization phase, only exploration is applied for a random time duration to initialize the Q-Table. In exploration, a value is randomly chosen for $(\rho,\Delta)$ from the available values of $\rho$ and $\Delta$. Thus, exploration helps find the impact of unexplored values of $\rho$ and $\Delta$, and consequently the Q-Table is enriched with new information. During exploitation, $(\rho,\Delta)$ that provides the maximum $Q(s_k,a_k)^{final}$ values in the Q-Table is selected. Therefore, the Q-Table serves as the experience from historical execution information for LIM2, where exploration helps enrich the Q-Table and exploitation helps enhance the Q-value. As a result, during the execution, LIM2 both can gather new information about an environment and exploit past knowledge. The probabilities of exploration and exploitation are $\epsilon_k$ and $(1-\epsilon_k)$, respectively, which are controlled by a random variable $\nu$. Hence, no explicit training and testing are followed by our model, and therefore LIM2 does not use any explicit dataset.

The proposed mechanism LIM2 runs at the gNB, i.e., one LIM2 module with one Q-Table per cell. However, a LIM2 module may have multiple RL agents depending on the number of handover operation initiations, where each handover is handled by one RL agent. When a new cell is formed by installing a gNB, one LIM2 module needs to be deployed in that cell. The functionality of a LIM2 module does not depend on the other ones, and consequently the RL agents of two different LIM2 module can work independently. As a result, if the number of cells is increased, the performance of LIM2 is not affected. Moreover, since each handover decision is executed by a separate RL agent, the individual handover operations are not impacted by the number of UEs in a cell. Therefore, the scalability of the service of LIM2 is not an issue.

Deep RL algorithms use deep learning to solve a problem, where a neural network (NN) is used to represent an RL policy, and very large inputs and a high volume of dataset are considered to train and test the NN. In LIM2, the RL agent considers only four inputs -- the RSRQ of neighbor cells, TTT, hysteresis margins, and Q-value, and thus LIM2 is characterized by a low overhead for handling input parameters. 
%The proposed LIM2 uses SARSA-based online learning and as an RL-based learning mechanism, SARSA does not require any predefined dataset, training, and testing. Thus, it is not required to train and test the proposed LIM2 model. 
However, in our context, a deep RL-based approach will be more complex in order to prepare a trained model. In addition, as per our knowledge, appropriate dataset is not available, that maintains our required input parameter set. LIM2 can adapt to different network conditions, which makes it scalable; whereas, a deep RL needs an explicit retraining with an updated dataset each time it is deployed in a new environment. Thus, there is also a scalability issue in the deep RL model. Hence, there is no advantage to move LIM2 to a deep RL-based implementation.

%\textcolor{blue}{The adaptability of LIM2 with different network conditions makes it scalable; whereas, a deep RL needs an explicit retraining by updated dataset each time when it is deployed in a new environment. Thus, there is also a scalability issue in a deep RL model. Hence, there will be no advantage of moving LIM2 to a deep RL-based implementation.}

Details of the execution steps of LIM2 are given in Algorithm~\ref{algo1} which follows three phases -- (i) selection of the target cell for a handover, (ii) selection of the TTT and hysteresis, and (iii) handover decision. These three phases are discussed in what follows.

\subsubsection{Target Cell Selection}

In this phase, the target cell and associated base station are selected based on the management report. Let the target base station and management report be $\mathcal{B}$ and $\mathcal{M}$, respectively. The RSRP values of the serving and target cells, and the RSRQ of the target cell are extracted from $\mathcal{M}$, and let these values be $U_{RSRP,k}^{srv}$, $U_{RSRP,k}^{nbr}$, and $V_{RSRQ,k}^{nbr}$, respectively. Then, using the Kalman filter, the serving cell evaluates $x_k^{srv}$ and $x_k^{nbr}$ of the neighbor cells. After that, for each of the neighbor cell, the serving cell calculates $Q(s_k,a_k)^{final}$ using (\ref{sarsa2}). Thus, the number of computed $Q(s_k,a_k)^{final}$ values is equal to the number of neighbor cells considered by the UE. The action $a_k$, that leads to the maximum $Q(s_k,a_k)^{final}$ value, is identified. This is because the objective is to maximize $Q(s_k,a_k)^{final}$ and choose the target cell such that the overall signal quality can be enhanced with a reduced noise. As a result, the throughput of the user will be increased. Assume that $\mathcal{T}_k$ is the target cell for action $a_k$. Let $x_{\mathcal{T}_k}^{nbr}$ be the a posteriori estimation of $U_{RSRP,k}^{nbr}$ of $\mathcal{T}_k$. After selecting the target cell using the $\epsilon$-greedy policy, the TTT and hysteresis values are chosen, as discussed next.

\subsubsection{TTT and Hysteresis Selection}

At the beginning of this phase, $\epsilon_k$ is calculated using (\ref{epsilon}). Based on a random function, the exploration/exploitation decision is determined, where $\epsilon_k$ controls the rate of exploration and exploitation. In exploitation, the pair $(\rho,\Delta)$, which provides the maximum Q-value ($Q(s_k,a_k)^{final}$) in the Q-Table, is chosen for the handover decision. As a result, the highest possible signal strength with the minimum environment noise will be experienced in the handover, maintaining a high throughput. During exploration, $(\rho,\Delta)$ is selected randomly since the learning agent needs to gain knowledge about the impact of a value of the pair $(\rho,\Delta)$ on the network performance. Thus, exploration helps gain experience about the performance of unexplored $(\rho,\Delta)$ values such that, based on this experience, the exploitation can enhance the system performance in the long run. $Q(s_k,a_k)^{final}$ is modified either after the exploration or exploitation. Hence, the Q-Table is updated after each of these $\epsilon$-greedy phases.  

When the proposed mechanism is initiated, the Q-Table is empty, and thus we need an initialization phase to populate the Q-Table. Therefore, in the initialization phase, only exploration is applied for a random time duration $t_{init}$ to initialize the Q-Table. 

\subsubsection{Handover Decision}

\begin{figure}[!t]
 \centering
 \includegraphics[width=\linewidth]{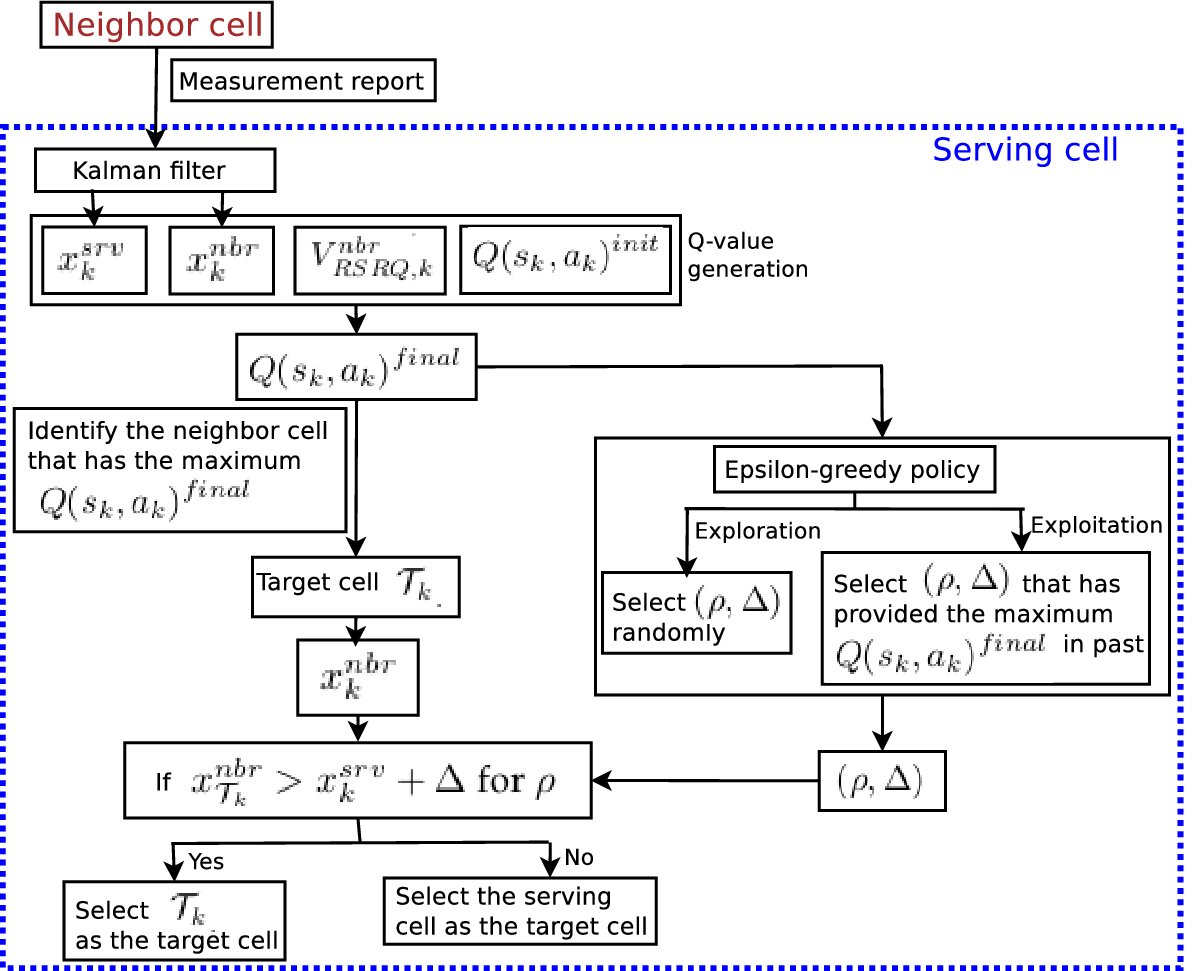}
 \caption{Overall mechanism of LIM2}
 \label{fig:model}
\end{figure}

After selecting the target cell and $(\rho,\Delta)$ for the handover decision, the handover triggering criteria is checked (we consider A3 event) as
\begin{equation}\label{event} 
 x_{\mathcal{T}_k}^{nbr} > x_k^{srv}+\Delta \qquad \text{for} \quad \rho.
\end{equation}
If the condition given in (\ref{event}) is satisfied for a duration of $\rho$, the handover decision is made and the base station of the target cell $\mathcal{T}_k$ becomes $\mathcal{B}$. However, if the condition is not satisfied, the handover is not performed, and the base station of the serving cell remains as $\mathcal{B}$. The overall mechanism of the proposed model is shown in Fig.~\ref{fig:model}.

\begin{algorithm}[!t]
\caption{LIM2 -- Algorithmic Description}
\label{algo1}
\begin{algorithmic}[1]
%\scriptsize
\State \textbf{Start}
\State \textbf{Input:} Measurement report $\mathcal{M}$. %containing posteriori estimations of $x_k^{srv}$ and $x_k^{nbr}$ by using Kalman filter.
\State \textbf{Output:} The target base station $\mathcal{B}$ for a handover. TTT and hysteresis.
%\State Calculate posteriori estimations of $x_k^{srv}$ and $x_k^{nbr}$ by using Kalman filter.
\State $U_{RSRP,k}^{srv} \leftarrow \text{getServingCellBeamPower}(\mathcal{M})$. %\Comment{selection of the target cell for a handover}
\State $U_{RSRP,k}^{nbr} \leftarrow \text{getNeighborCellBeamPower}(\mathcal{M})$.
\State $V_{RSRQ,k}^{nbr} \leftarrow \text{getNeighborCellBeamPower}(\mathcal{M})$.
\State $x_k^{srv} \leftarrow \text{KalmanFilter}(U_{RSRP,k}^{srv})$.
\State $x_k^{nbr} \leftarrow \text{KalmanFilter}(U_{RSRP,k}^{nbr})$.
%\State Calculate posteriori estimations of $x_k^{srv}$ and $x_k^{nbr}$ by using Kalman filter.
\State $Q(s_k,a_k)^{final} \leftarrow Q(s_k,a_k)^{init}+\alpha[V_{RSRQ,k}^{nbr}+\gamma x_k^{nbr}-x_k^{srv}]$.
%\State $Q(s_k,a_k)_{max}^{final} \leftarrow \text{max}\{Q(s_k,a_k)^{final}\}$.
\State $a_k \leftarrow \underset{a}{\argmax}\{Q(s_k,a)^{final}\}$.
%\State $\mathcal{T} \leftarrow getTargetCell\big(Q(s_k,a_k)_{max}^{final}\big)$.
\State $\mathcal{T}_k \leftarrow getTargetCell(a_k)$.
\State $x_{\mathcal{T}_k}^{nbr} \leftarrow getX(\mathcal{T}_k)$.
%\State $\rho \leftarrow \text{getTimeToTrigger}()$.
\State $\epsilon_k \leftarrow \text{calculateEpsilon()}$.
\State Let $\nu \leftarrow $ Random(0,1).
\If{$\nu \leq \epsilon_k$}
    \State Choose $(\rho,\Delta)$ which provides the maximum Q-value ($Q(s_k,a_k)^{final}$) in the Q-Table.
\Else
  \State Choose $(\rho,\Delta)$ randomly.
\EndIf
\If{$x_{\mathcal{T}_k}^{nbr} > x_k^{srv}+\Delta$ for $\rho$} 
  \State $\mathcal{B} \leftarrow \text{getBaseStation}(\mathcal{T}_k)$.
\Else 
  \State $\mathcal{B} \leftarrow \text{getServingCellBaseStation}()$.
\EndIf
\State Return $\mathcal{B}$.
\State \textbf{End}
\end{algorithmic}
\end{algorithm}

\section{Performance Analysis}
\label{sec:impl}

We analyze the performance of LIM2 by implementing it in NS-3.33~\cite{ns3}, where the 5G NR module is plugged into \texttt{ns-3-dev}. %The 5G NR of frequency range $2$ (FR2) (mmWave) is considered in our implementation and it supports 3GPP-complying buffer status reporting. LIM2 runs at the gNB.
The number of gNBs is set to $50$, and thus $50$ cells are formed in the mobile network. 
Particularly, in our model, the action is the ID of the target cell for a handover. Therefore, the cardinality of the action space depends on the maximum number of available cells in a network. Since we consider $50$ cells in the performance analysis of the proposed mechanism, the cardinality of the action space is $50$ in our model.
At the beginning of the simulation, $10$ UEs are placed in each cell following a Poisson distribution centered at the gNB's position. The SNR value is selected randomly between $20$dB-$45$dB. Each simulation instance is run for $100$ times, where each run has a duration of $2$s. Thus, results are shown as an average of $100$ runs of a simulation instance containing both uplink and downlink transmissions. The speed of the UEs is represented in km/h and chosen from the set $\{50,100,150,200,250,300,350\}$. The propagation and shadowing effects are computed through the \texttt{MmWave3gppPropagationLossModel}. In the simulation, both UDP and TCP traffic are considered representing $60\%$ and $40\%$ of the total traffic, respectively. We compute the UDP and TCP throughput to analyze the performance in terms of average throughput. 

In Algorithm~\ref{algo1}, the values of $U_{RSRP,k}^{srv}$, $U_{RSRP,k}^{nbr}$, $V_{RSRQ,k}^{nbr}$, $x_k^{srv}$, $x_k^{nbr}$, and $\epsilon_k$ are dynamically set during the execution of the algorithm. Based on these values, $Q(s_k,a_k)^{final}$, $a_k$, $\mathcal{T}_k$, and $x_{\mathcal{T}_k}^{nbr}$ are computed at the run time of the algorithm. However, $\alpha$, $\gamma$, $N$, and $r$ are set to fixed values before starting the execution of the algorithm, and therefore these parameters impact the performance of the proposed mechanism. We set $\alpha=0.1$, $\gamma=0.5$, $N=47$ (sum of the number of TTT and hysteresis margins), and $r=1.0$. %Moreover, two ranges of values are assigned to the hysteresis mergin $\Delta$, and TTT $\rho$ separately, where $\Delta$ varies from $0$ dB to $30$ dB, separated by $1.0$ dB and $\rho=\{0,40,64,80,100,128,160,256,320,480,512,640,1024,1280,2560,5120\}$ms.
In SARSA, $\alpha$ is set to $0.1$ such that the learning agent can be assigned a significant time to obtain information about the environment. To impose a balance between current and future reward values, $\gamma$ is set to $0.5$. 
In the performance analysis, unless specified otherwise, the speed of the UEs is $200$km/h, and this speed is considered to impose high mobility (such as high speed train)~\cite{li2020beyond}. Details of the simulation parameters are given in Table~\ref{table:sim}.

\noindent{\textbf{Handover triggering event, the TTT, and hysteresis margin:}}
We consider event A3 as the handover triggering criteria. In the exploration phase, the hysteresis margin is chosen between $0$ and $30$ dB, where two margins are separated by $1$ dB. Therefore, a total of $31$ values are available for the hysteresis margin. The TTT is selected from the set $\{0,40,64,80,100,128,160,256,320,480,512,640,1024,1280,\\2560,5120\}$ms~\cite{3GPP1,3GPP2}.
\begin{table}[!t]
\caption{Simulation Parameters}
%\scriptsize
\centering
\begin{tabular}{|p{3.2cm}|p{4.5cm}|}
\hline
\textbf{Parameter} & \textbf{Value}\\
\hline
Cell radious & $150$ m \\
\hline
Center frequency & $26$ GHz\\
\hline
TxPower of UE & $23$ dBm \\
\hline
TxPower of gNB & $46$ dBm \\
\hline
Speed of UE & Constant velocity mobility model \\ 
\hline
Simulation time & $2$s\\
\hline
Data rate in Evolved Packet Core (EPC) & $100$ Gbps\\
\hline
Channel bandwidth & $400$ MHz\\
\hline
Mobility model & Constant velocity mobility model\\
\hline
Path loss model & Log-normal path loss model (path loss exponent=$3.0$)\\
\hline
Propagation delay model & Constant speed propagation delay model\\
\hline
Fading Model & Friis spectrum propagation loss model\\
\hline
Bit error rate (BER) & $0.03$ \\
\hline
Adaptive Modulation and Coding (AMC) model & Vienna \\
\hline
UE scheduler type & PfFfMacScheduler \\
\hline 
NoiseFigure of UE & $9$ \\
\hline
NoiseFigure of gNB & $5$ \\
\hline
DefaultTransmissionMode & $0$ (SISO)\\
\hline
Antenna pattern & Omnidirectional\\
\hline
Thermal noise density & $-174$ dBm/Hz\\
\hline
Value of $N$ & $47$ (sum of the number of the TTT and hysteresis)\\
\hline
Value of $r$ & $1.0$\\
\hline
Values of $\alpha$ and $\gamma$ & $\alpha = 0.1$, $\gamma = 0.5$\\
\hline
Value of $\rho$ & $\rho$ varies from $0$ dB to $30$ dB, separated by $1.0$ dB\\
\hline
%\textcolor{blue}{Value of $\Delta$} & \textcolor{blue}$\{0,40,64,80,100,128,160,256,320,480,512,640,1024,1280,2560,5120\}$}\\
%\hline
\end{tabular} 
\label{table:sim}
\end{table}

\subsection{Baseline Mechanisms}

To analyze the performance of LIM2, we use Reliable Extreme Mobility (REM)~\cite{li2020beyond} and Contextual Multi-Armed Bandit (CMAB)~\cite{yajnanarayana20205g} as baseline mechanisms. REM is based on the delay-Doppler domain and considers movement-based mobility management in 5G and beyond. In the delay-Doppler domain, REM uses a signaling overlay that extracts the client's movement pattern and multi-path outline with the orthogonal time-frequency space (OTFS) modulation, where the handover is performed based on the extracted client profile. To stabilize the signaling, REM uses a scheduling-based OTFS. On the other hand, CMAB applies a reinforcement learning mechanism to perform the handover in 5G networks, where a centralized agent is designed to select appropriate handover actions based on measurement reports from UEs. In CMAB, the goal is to choose the target cell such that the throughput is maximized after the migration to the new cell. However, both REM and CMAB do not dynamically adjust the TTT and hysteresis in the handover execution.

\subsection{Average Throughput and Packet Loss Rate}

\begin{figure}[!t]
 \centering
 \includegraphics[width=\linewidth]{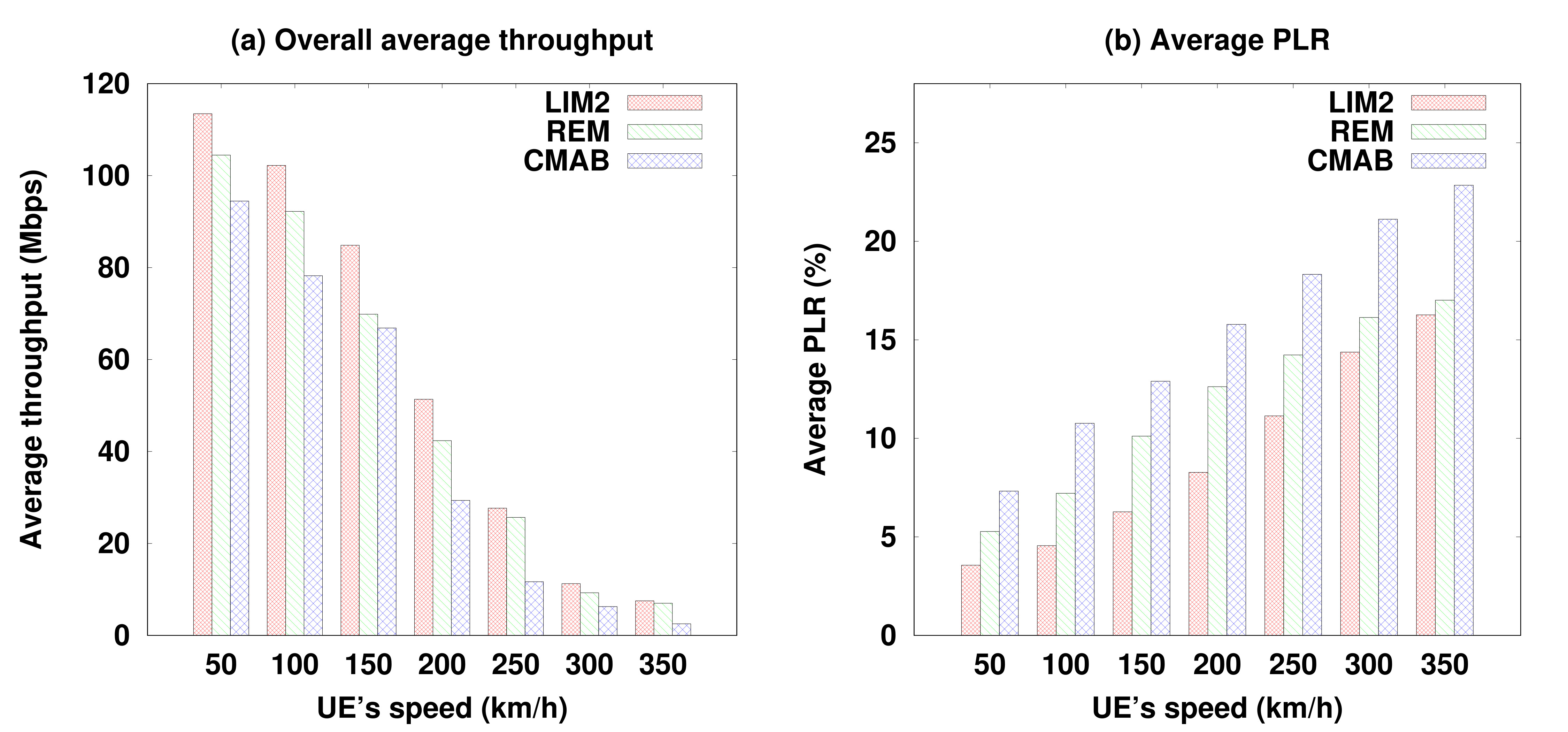}
 \caption{(a) Average throughput and (b) Average packet loss rate}
 \label{fig:thrplr}
 %\vspace{-5mm}
\end{figure}

\noindent{\textbf{Throughput:}} Fig.~\ref{fig:thrplr}(a) shows the average throughput of LIM2 and other baselines. In LIM2, the action $a_k$ is chosen to maximize the value of $Q(s_k,a_k)^{final}$, which leads to the reduction of the noise and increase in the overall signal strength in the cell. The improvement of the signal quality results in an increase in average throughput in the network. %To calculate $Q(s_k,a_k)^{final}$, the computation of the a posteriori of the RSRP (using the Kalman filter) helps select the target cell by considering future signal qualities of the serving and neighbor cells. %Since the target cell is selected based on the maximization of $Q(s_k,a_k)^{final}$, the average throughput is not affected after handover. 
In this context, the target cell is chosen intelligently such that high RSRP and RSRQ are maintained to achieve a high throughput under mobility after handover. Moreover, LIM2 is an online learning-based scheme that adjusts the TTT and hysteresis based on $Q(s_k,a_k)^{final}$. As a result, the handover time is intelligently controlled to maintain a high throughput under mobility.
On the other hand, REM is not an online learning-based approach and does not consider the RSRP, RSRQ, and noise, which are important factors for selecting a target cell and maintaining a high throughput under mobility. %These factors become more crucial when the target cell needs to be chosen from a set of neighbor cells and the speed of a mobile device increases. 
%In this scenario, the target cell should be chosen intelligently such that high RSRP and RSRQ are maintained to achieve a high throughput under mobility after handover. Moreover, since LIM2 is an online learning-based scheme, it has the tendency to adjust TTT and hysteresis based on $Q(s_k,a_k)^{final}$. As a result, the handover time is intelligently controlled to maintain a high throughput under mobility. 

Although CMAB applies reinforcement learning, the handover is performed based on the present RSRP value, and static TTT and hysteresis. From Fig.~\ref{fig:thrplr}(a), it is noted that LIM2 has a significantly higher average throughput than CMAB. When the UE's speed is lower than $200$km/h, LIM2 also achieves a significantly higher throughput than REM. Thanks to the TTT and hysteresis, LIM2 provides a comparatively higher average throughput than REM when the UE's speed is more than $200$ km/h.     
From Fig.~\ref{fig:thrplr}(a), LIM2 has approximately $19\%$ and $68\%$ higher average throughputs than the REM and CMAB schemes, respectively. 

\noindent{\textbf{Packet loss rate:}} We compute the packet loss rate (PLR), which is the ratio of the number of lost packets to the number of successfully transmitted packets during a session.
Since LIM2 predicts the future signal quality of neighbor cells, the selected target cell would have a higher RSRP than the neighbor cells after the handover, and consequently the PLR is significantly reduced in the target cell, as shown in Fig.~\ref{fig:thrplr}(b). In addition, due to the intelligent adjustment of the TTT and hysteresis, the probability of maintaining the appropriate timing in handover execution is higher for LIM2 than other baselines, which also reduces the overall PLR.
From Fig.~\ref{fig:thrplr}(b), LIM2 has an average PLR approximately $28\%$ and $42\%$ lower than REM and CMAB, respectively.

\subsection{Throughput, PLR, Packet Delay, and Handover Latency With Respect to Cell Crossing Rate}

\begin{figure}[!t]
 \centering
 \includegraphics[width=\linewidth]{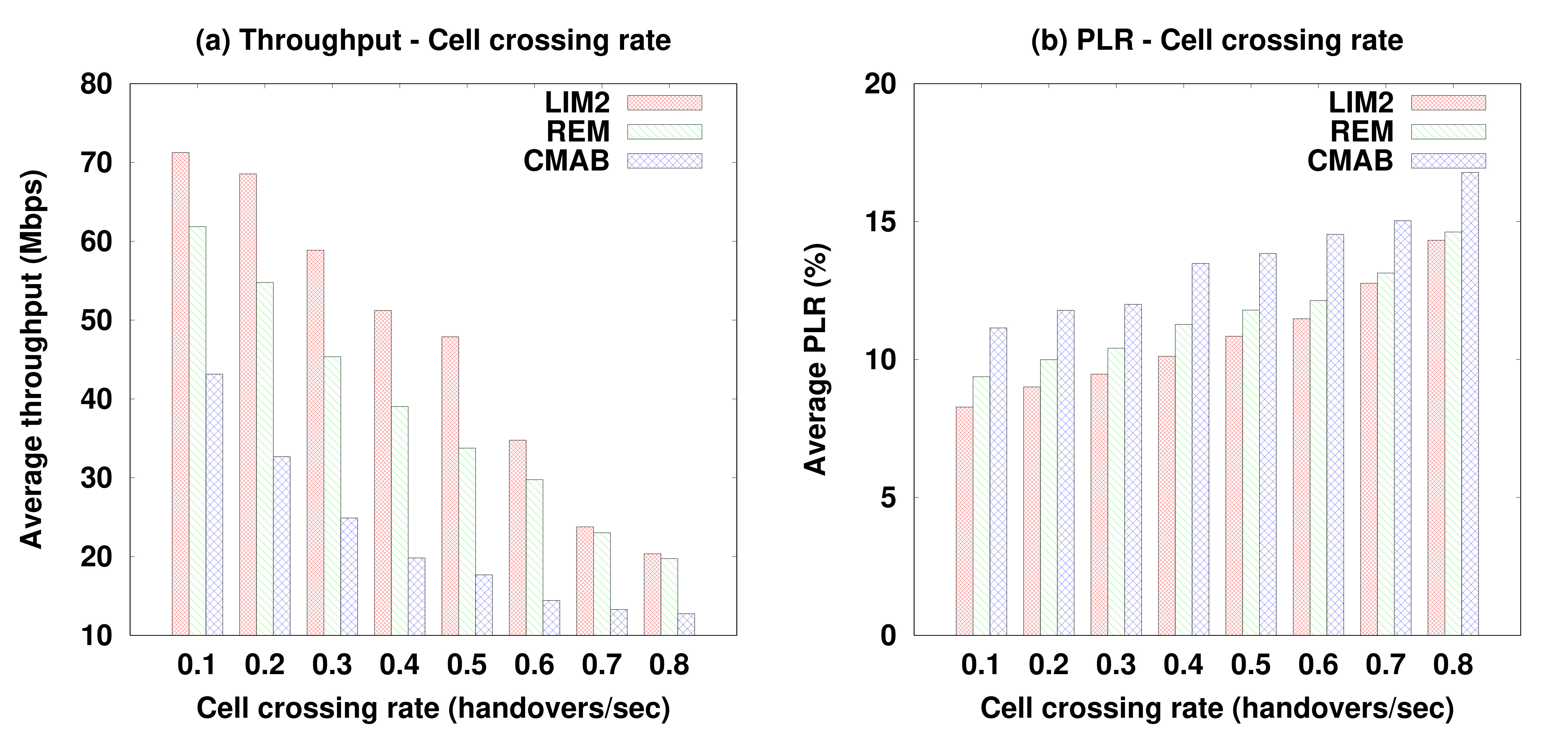}
 \caption{Analysis of average throughput and packet loss rate with respect to cell crossing rate}
 \label{fig:thrplrcell}
 %\vspace{-5mm}
\end{figure}
\begin{figure}[!t]
 \centering
 \includegraphics[width=\linewidth]{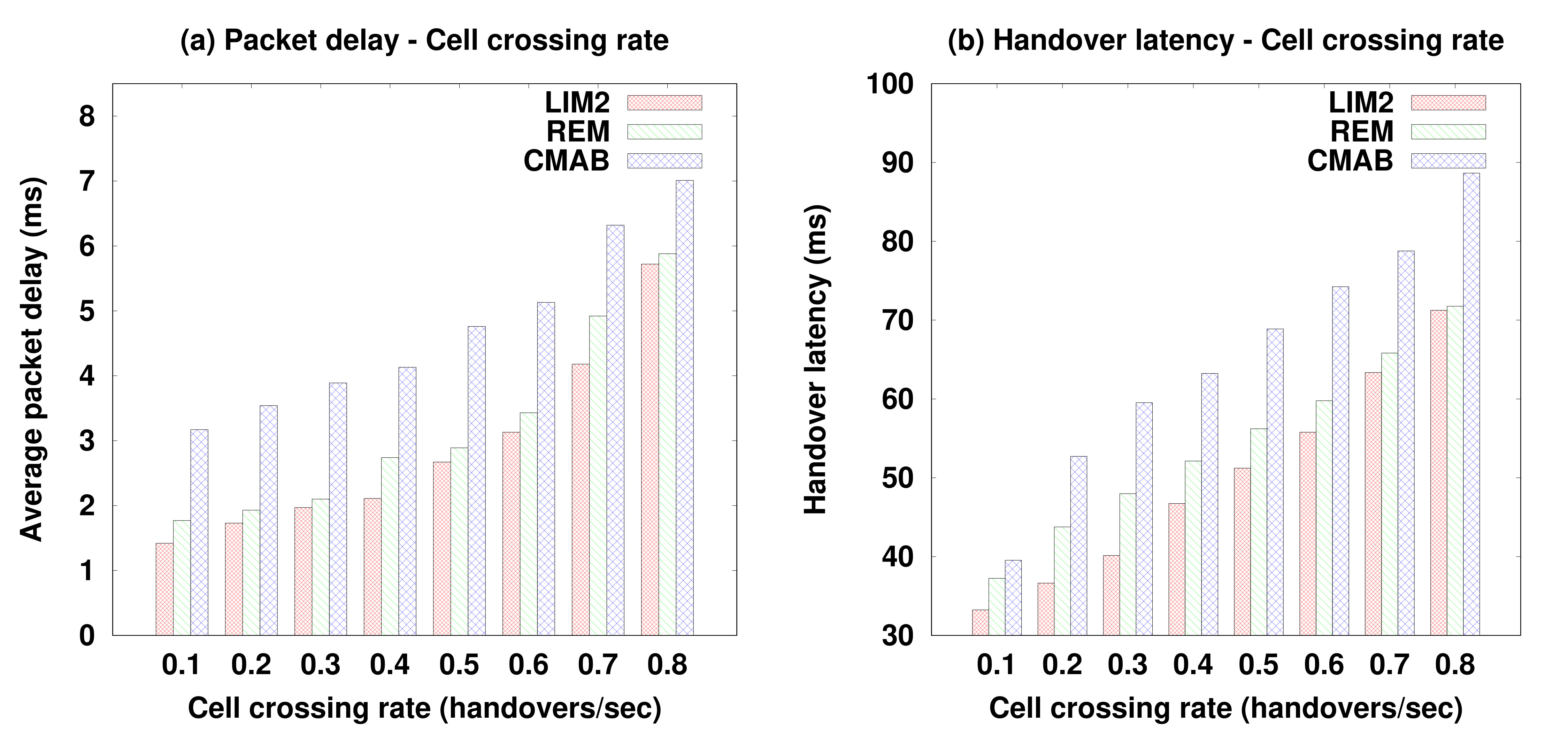}
 \caption{Analysis of average packet delay and handover latency with respect to cell crossing rate}
 \label{fig:delaycell}
 %\vspace{-5mm}
\end{figure}

\noindent{\textbf{Throughput and PLR:}} Fig.~\ref{fig:thrplrcell} shows the average throughput and PLR with respect to the cell crossing rate, where it is noted that LIM2 has a higher average throughput and lower PLR than other baselines. The TTT and hysteresis significantly impact the handover execution time. A higher TTT and hysteresis delays the transition from the serving cell to the target cell. However, lower values of the TTT and hysteresis may cause ping-pong effects in handover. LIM2 tries to adapt the TTT and hysteresis by considering the future signal strength of the serving and target cells, and the past knowledge of the adjustment of these parameters. As a result, an average high throughput can be maintained during the handover procedure. Therefore, LIM2 provides a higher average throughput against cell crossing rate than REM and CMAB.
If the handover is not completed within the required time, a significant volume of packets will be dropped in the target cell. LIM2 addresses this problem by dynamically selecting the TTT and hysteresis, where the consideration of the maximization of $Q(s_k,a_k)^{final}$ leads to a transition to the target cell such that the handover is executed at an optimal time.

\noindent{\textbf{Packet delay and handover latency:}} The packet delay is the time interval required to transmit a packet from the source to the destination. The handover latency is the time period between the reception (or transmission) of the last packet through the old cell connection and the first packet in the new cell connection. The handover latency depends on the handover initialization, decision, and execution.
Fig.~\ref{fig:delaycell} shows the average packet transmission delay and handover latency with respect to the number of handovers per second, where LIM2 provides a lower packet delay and handover latency than baselines. Due to the intelligent adjustment of the TTT and hysteresis, the probability of maintaining the appropriate delay in handover execution is higher in LIM2 than other baselines, and consequently the handover latency and packet transmission delay are reduced. In this regard, the Q-Table serves as past experience on the adjustment of the TTT and hysteresis, which helps optimize these two parameters in future handover executions. In REM and CMAB approaches, the threshold-based adjustment of the TTT and hysteresis fails to tune the handover latency based on the present scenario. %, which is influenced further by the high mobility of the mobile devices. 

Table~\ref{table1} presents a comparative analysis of the throughput, PLR, packet delay, and handover latency, with respect to the cell crossing rate, as shown in Figs.~\ref{fig:thrplrcell} and~\ref{fig:delaycell}.   
\begin{table}[h]
\caption{Performance Gain of LIM2 With Respect To Cell Crossing Rate}
\scriptsize
\centering
\begin{tabular}{|p{1.3cm}|p{1.3cm}|p{1.25cm}|p{1.4cm}|p{1.3cm}|}
\hline
\textbf{Scheme} & \textbf{Throughput gain (\%)} & \textbf{PLR reduction (\%)} & \textbf{Packet delay reduction (\%)} & \textbf{Handover latency reduction (\%)}\\
\hline
\hline
With respect to REM & $16.83\%$ higher & $5.44\%$ lower & $8.74\%$ lower & $6.69\%$ lower\\  
\hline 
With respect to CMAB & $87.31\%$ higher & $21.12\%$ lower & $38.98\%$ lower & $24.87\%$ lower\\ 
\hline
\end{tabular} 
\label{table1}
\end{table}

\subsection{Block Error Rate under Different SNR Values}

\begin{figure}[!t]
 \centering
 \includegraphics[width=\linewidth]{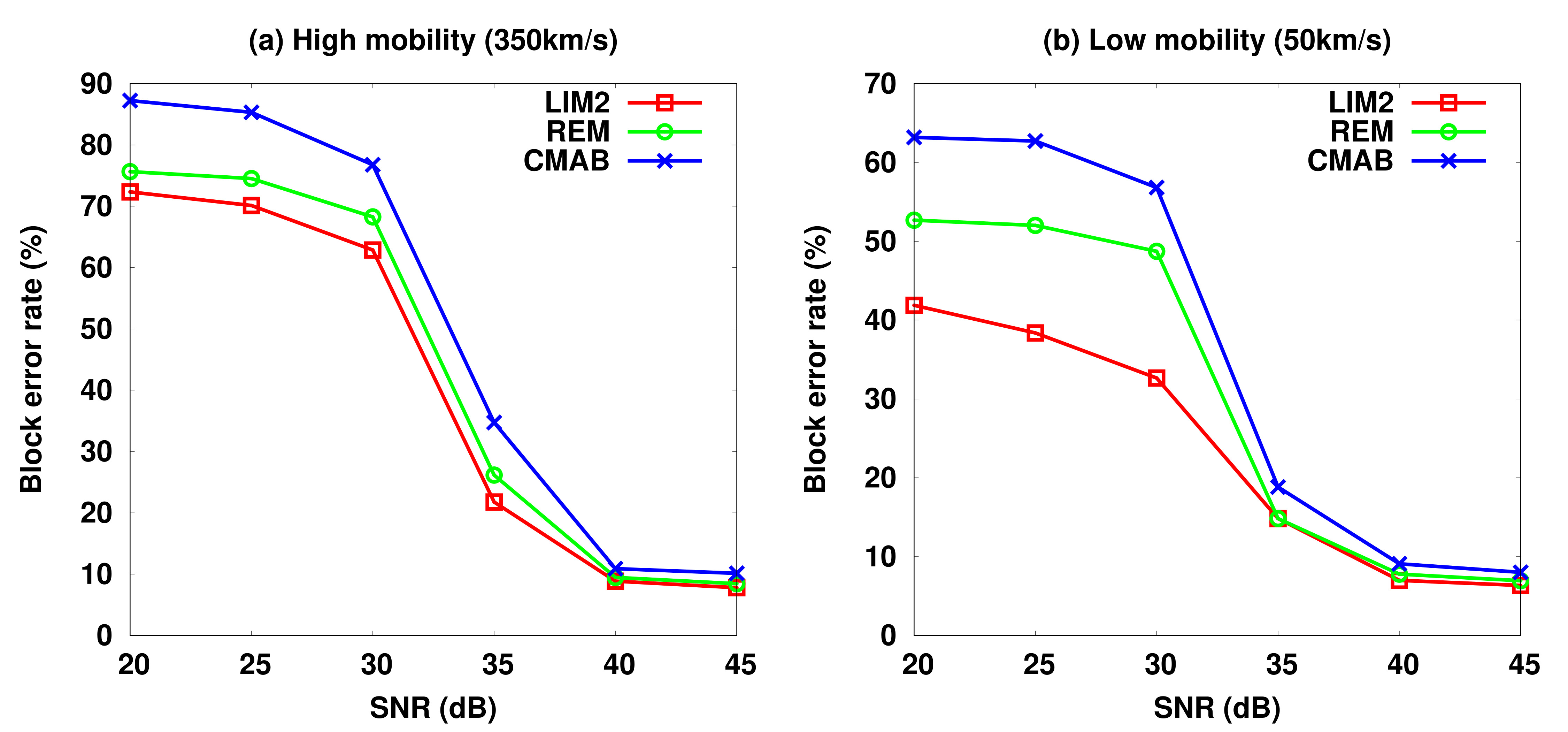}
 \caption{Block error rate under different SNR values for (a) high mobility and (b) low mobility}
 \label{fig:BLR}
 %\vspace{-5mm}
\end{figure}

In 5G NR, the transport block is the payload which is transferred between the medium access control (MAC) and physical (PHY) layers. Fig.~\ref{fig:BLR} shows the transport block error rate under different SNR values and mobility conditions, where it is observed that, when the SNR is less than $30$dB, the block error rate for LIM2 is significantly lower than for REM and CMAB. However, as the SNR increases, there is no notable difference in the performance of LIM2 and REM in terms of block error rate. Therefore, LIM2 has better adaptability in low signal strengths because it migrates to the target cell based on the prediction of the RSRP in the next timestamp. Considering the RSRP, the maximization of $Q(s_k,a_k)^{final}$ helps reduce the block error rate in the target cell. As the mobility increases, the block error rate increases (Fig.~\ref{fig:BLR}(a)); however, in LIM2, the consideration of the RSRQ along with the predicted RSRP helps choose the target cell with higher signal quality, leading to a reduction of the block error rate. 
In the high mobility scenario, when the SNR is $25$dB, LIM2 has a block error rate approximately $6\%$ and $16\%$ lower than the REM and CMAB schemes, respectively. However, in the low mobility case, the block error rate of LIM2 is approximately $24\%$ and $36\%$ lower than REM and CMAB, respectively. 

\subsection{Distribution of Handover Failure and Average Throughput}

\begin{figure}[!t]
 \centering
 \includegraphics[width=\linewidth]{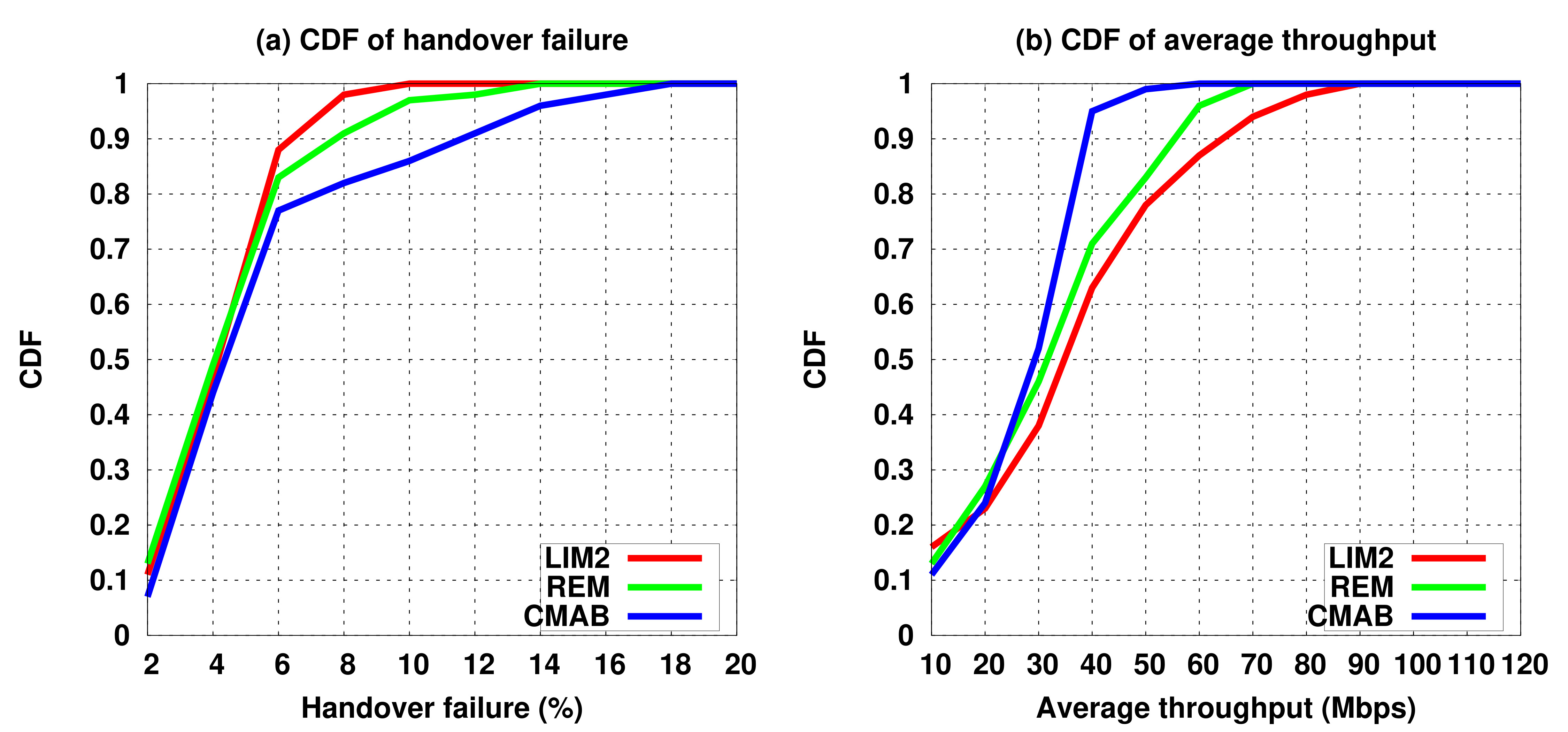}
 \caption{(a) Distribution of handover failure and (b) distribution of average throughput}
 \label{fig:cdf}
 %\vspace{-5mm}
\end{figure}

The cumulative density functions (CDFs) of the handover failure rate and average throughput are shown in Fig.~\ref{fig:cdf}, wherein the speed of the UEs is set to $200$km/h. From Fig.~\ref{fig:cdf}(a), it is noted that the distribution of the handover failure in LIM2 is high when the handover failure rate is in the $6\%$-$10\%$ range; whereas, in REM, the distribution is high when the handover failure rate lies in the $6\%$-$14\%$ range. However, CMAB has a significant higher CDF of handover failure (the handover failure rate is up to $18\%$) than LIM2 and REM. Since both the RSRP and RSRQ of neighbor cells are considered to perform the handover, the signal interference is also included with the received power to ensure an appropriate selection of the target cell for the handover. In addition, the online learning-based adaptation of the TTT and hysteresis leads to a suitable setting of these parameters based on the computed $Q(s_k,a_k)^{final}$ value. Consequently, the handover is executed by concentrating on both the serving and neighbor cells' overall signal quality which helps increase the probability of correct handover decisions. As a result, LIM2 is also able to achieve a higher distribution of average throughput than REM and CMAB, as shown in Fig.~\ref{fig:cdf}(b). From this figure, it is noted that the average throughput distribution in LIM2 is between $30$-$80$ Mbps; whereas, the distributions in REM and CMAB are bounded by $70$ Mbps and $50$ Mbps, respectively. 
% which indicates that $6\%$-$10\%$ of the handover failure has the maximum probability of the occurrence

\subsection{Handover Failure With Respect to Change in SNR and Speed of Mobile Devices}

\begin{figure}[!t]
 \centering
 \includegraphics[width=\linewidth]{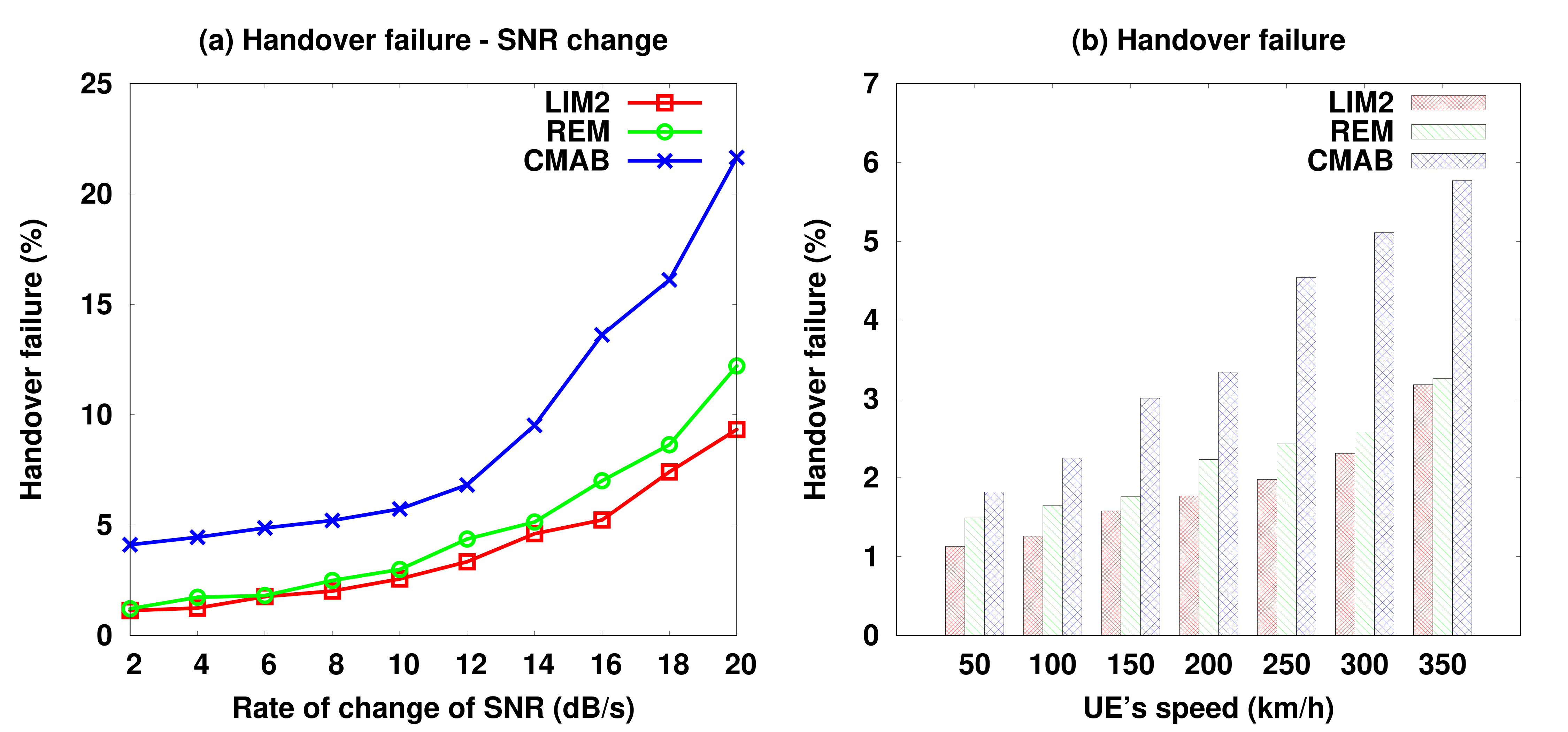}
 \caption{Handover failure against the rate of change in SNR and the UE's speed}
 \label{fig:HOF}
 %\vspace{-5mm}
\end{figure}

The handover failure rates against the change in SNR and speed of UEs are shown in Fig.~\ref{fig:HOF}, where LIM2 shows lower handover failures than REM and CMAB. At a low rate of change in SNR, REM has a slightly higher handover failure rate than LIM2; however, as the rate of change in SNR increases, the handover failure rate becomes higher in REM than for LIM2, as shown in Fig.~\ref{fig:HOF}(a). Due to the signal quality based dynamic adaptation of the TTT and hysteresis, LIM2 is able to intelligently execute the handover by considering the signal strength variation in the network. Moreover, under high mobility, a mobile device also experiences a frequent change in the channel's signal strength. Since LIM2 can adapt better to changes in the signal quality than baseline schemes, LIM2 provides lower handover failures than REM and CMAB under different UE speeds. Under very high speed ($350$ km/h) scenarios, LIM2 also has a slightly lower handover failure than REM, as shown in Fig.~\ref{fig:HOF}(b). In the high mobility case, LIM2 has a handover failure rate approximately $2\%$ and $44\%$ lower than REM and CMAB, respectively. 

\subsection{Convergence Analysis of LIM2}

\begin{figure}[!t]
\centering
\includegraphics[width=\linewidth]{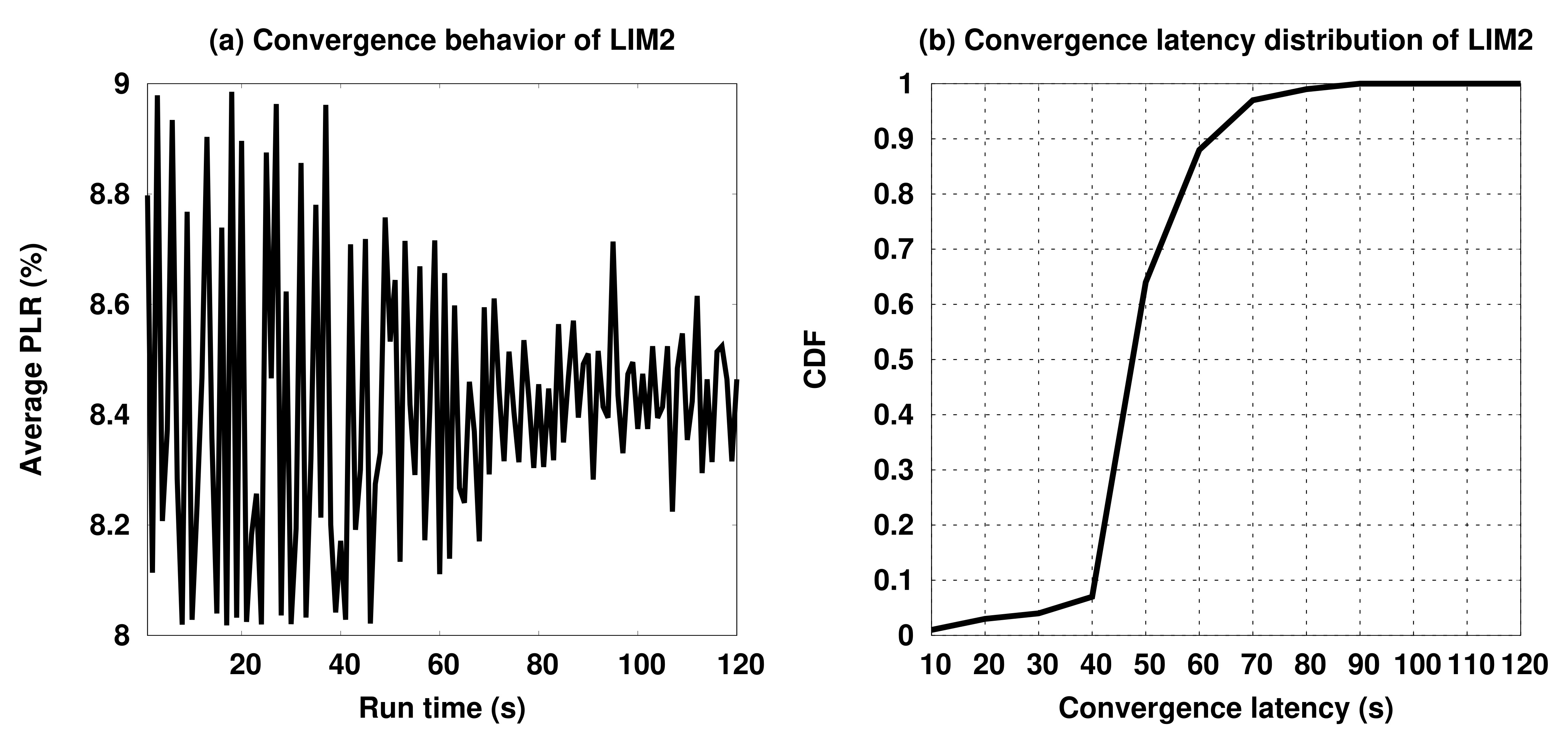}
%\vspace{-30mm}
\caption{(a) Convergence behavior and (b) distribution of convergence latency}
\label{fig:conv}
%\vspace{-5mm}
\end{figure}

In order to analyze the convergence behavior of LIM2, we calculate the average PLR at each timestamp of the simulation, where the timestamp is set to $1$s. At the early stage of the execution, the exploration is applied frequently to obtain information about the TTT and hysteresis values, and therefore a significant fluctuation in the average PLR is observed in initial timestamps, as shown in Fig.~\ref{fig:conv}(a). As time increases, the volume of the Q-Table is increased, which helps exploit past knowledge for the adaptation of the TTT and hysteresis. As a result, LIM2 provides quite stable average PLR as the run time progresses, compared to the beginning of the execution. From Fig.~\ref{fig:conv}(a), it is noted that LIM2 converges to a comparatively low fluctuation at approximately $70$s which is quite low considering the handover requirements in high speed mobility.

We run the convergence analysis experiment $50$ times and compute the distribution of the convergence latency, shown in Fig.~\ref{fig:conv}(b). From this figure, it is noted that the CDF is high between $40$--$80$s. If a similar network condition is observed in the past, the convergence is reached quickly. However, since the exploration selects values randomly, the learning converges when an optimal solution is observed.

\subsection{Reliability Analysis}

\begin{figure}[!t]
\centering
\includegraphics[width=\linewidth]{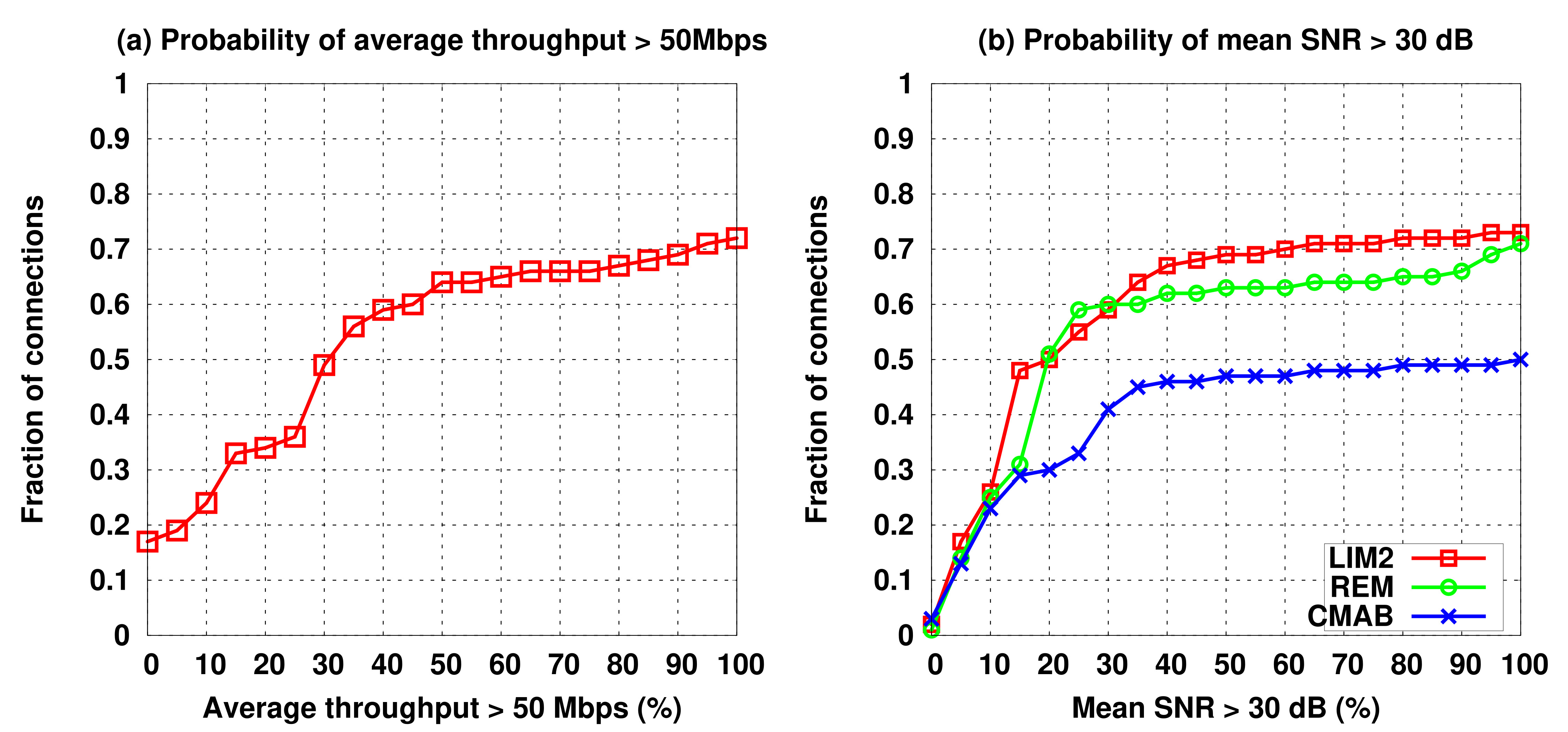}
%\vspace{-30mm}
\caption{(a) Fraction of connections of achieving greater than $50$ Mbps throughput in LIM2 and (b) Fraction of connections having greater than $30$ dB mean SNR}
\label{fig:rel}
%\vspace{-5mm}
\end{figure}

\begin{figure}[!t]
\centering
\includegraphics[width=\linewidth]{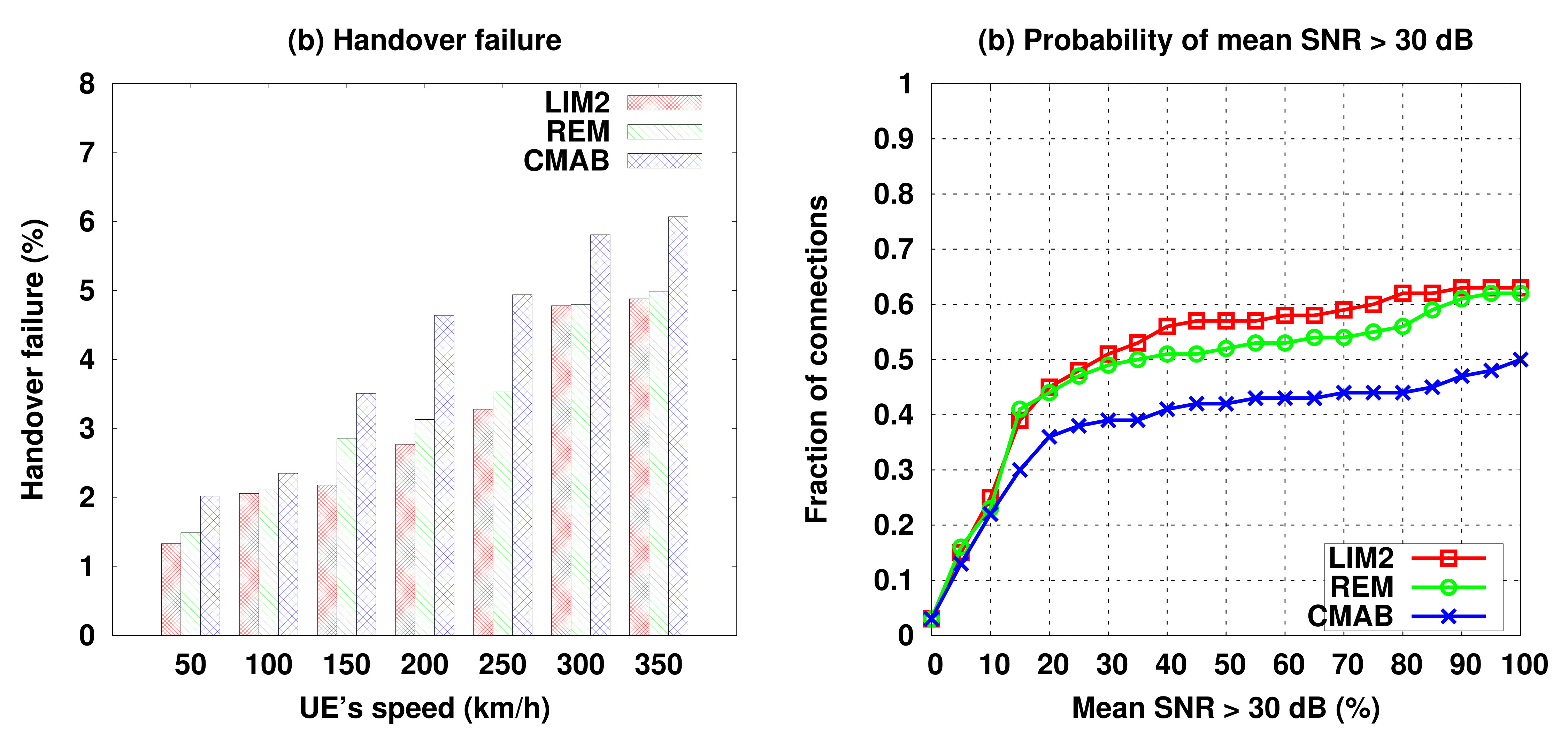}
%\vspace{-30mm}
\caption{(a) Handover failure against different UE's speed and (b) Fraction of connections having greater than $30$ dB mean SNR}
\label{fig:realistic}
%\vspace{-5mm}
\end{figure}

Fig.~\ref{fig:rel} and Fig.~\ref{fig:realistic} show the reliability analysis of LIM2. %The figure helps analyze the reliability in maintaining high throughput after handover and the reliability of the signal quality prediction. 
For each connection after handover, Fig.~\ref{fig:rel}(a) and Fig.~\ref{fig:rel}(b) show the fraction of executions where the achieved average throughput is greater than $50$ Mbps and the mean SNR is greater than $30$ dB, respectively. In Fig.~\ref{fig:rel}(a), we consider LIM2 only and set the average throughput threshold to $50$ Mbps because when the speed of the UEs is $200$ km/h, only LIM2 can reach an average throughput of more than $50$ Mbps (Fig. 7(a)). %For the mean SNR, $30$ dB can be considered as a minimum average requirement of SNR to ensure that the block error rate is not significantly high (Fig.~\ref{fig:BLR}). 
In Fig.~\ref{fig:rel}(b), we consider varying SNRs selected randomly between $20$dB-$45$dB. The simulatiom is run for $100$ times. Then, considering all the runs, we capture the probability of the connection that experiences a mean SNR (over time) greater than 30 dB. In this context, we consider $11$ probability values of the mean SNR, as shown in Fig.~\ref{fig:rel}(b). For the mean SNR, $30$ dB can be considered as a minimum average SNR requirement, where the highest block error rate is approximately $76\%$ (Fig.~\ref{fig:BLR}) and it is occurred for CMAB in the high mobility scenario. However, if the SNR is decreased further, the block error rate exceeds $85\%$ for CMAB.
%In Fig.~\ref{fig:rel}(b), we consider varying SNRs and the SNR values are selected randomly between $20$dB-$45$dB. The simulatiom is run for $100$ times. Then, we capture the fraction of connections when the probability of the mean SNR is greater than $30$ dB considering all the runs. In this context, we consider $11$ probability values of the mean SNR, as shown in Fig.~\ref{fig:rel}(b).}
From Fig.~\ref{fig:rel}(a), it can be noted that after handover, most of the connections achieve an average throughput greater than $50$ Mbps, where there is a probability of 0.8 that $67\%$ of a connection duration maintains an average throughput greater than $50$ Mbps. Due to the adaptive selection of the TTT and hysteresis margins based on the integration of the RSRP and RSRQ, LIM2 is able to maintain an average throughput greater than $50$ Mbps with a higher probability in most of the time in a connection after handover. From Fig.~\ref{fig:rel}(b), we can observe that in LIM2, there is an $80\%$ probability that $72\%$ of the connection duration experiences a mean SNR (over time) greater than $30$ dB. This is due to the Kalman filter-based prediction of the RSRP to choose the target cell. In this context, SARSA plays a crucial role by integrating the RSRP with the RSRQ and maximizing the Q-value, which helps ensure a higher signal strength is obtained for most of the time after the connection with the selected target cell. However, due to the lack of adaptability, in REM and CMAB, there is an $80\%$ probability that $65\%$ and $49\%$ of the connection duration experience a mean SNR (over time) greater than $30$ dB, respectively. Therefore, LIM2 is more reliable for signal quality predictions than other baselines.

To include more realistic channel settings for high-speed train, we consider the \texttt{ThreeGppV2vHighwayPropagationLossModel}, \texttt{RandomPropagationDelayModel}, and \texttt{Nakagami}-\texttt{PropagationLossModel} as propagation loss, propagation delay, and fading models, respectively. The results are shown in Fig.~\ref{fig:realistic}. In high speed rail, the dopper effect is high. Since the \texttt{ThreeGppV2vHighwayPropagationLossModel} provides high doppler effect, we use this model as propagation loss model in our implementation, where the doppler frequency is set to $1300$ Hz.
In \texttt{Random}-\texttt{PropagationDelayModel}, the propagation delay between every pair of nodes is totally random. Moreover, the delay is different for each packet sent in the network. \texttt{NakagamiPropagationLossModel} considers high variations in signal strength, which occurs due to multipath fading.

The handover failure rates against the change in speed of UEs are shown in Fig.~\ref{fig:realistic}(a), where it is noted that LIM2 shows lower handover failures than REM and CMAB, even in more realistic channel conditions. In this case, under very high speed (greater than $250$ km/h) scenarios, LIM2 also has a marginally lower handover failure than REM, which is due to its tendency of predicting and learning the channel condition before handover. From Fig.~\ref{fig:realistic}(b), we can observe that LIM2 is also more reliable for signal quality predictions than baselines, and that better reliability in LIM2 is achieved when considering both RSRP and RSRQ. In frequently changing channel conditions, the online learning-based adaptation of the TTT and hysteresis helps LIM2 perform handover considering the best promising target cell that can have the high probability of experiencing higher signal strength than neighbor cells. From Fig.~\ref{fig:realistic}(b), in LIM2, there is an $80\%$ probability that $62\%$ of the connection duration experiences a mean SNR (over time) greater than $30$ dB. Whereas, in REM and CMAB, there is an $80\%$ probability that $56\%$ and $44\%$ of the connection duration experience a mean SNR (over time) greater than $30$ dB, respectively.

\section{Conclusion}
\label{sec:concl}

High speed mobility management is a great challenge in 5G and beyond technologies. In this direction, we propose an online learning-based mechanism, namely LIM2. Using a Kalman filter, LIM2 computes the a posteriori of the RSRP values of the serving and neighbor cells to identify the best target cell for handover, such that after the migration, high performance is maintained in extreme mobility. Based on the estimated signal quality, a SARSA-based selection of the target cell makes LIM2 an intelligent approach, leading to a dynamic handover decision considering future network conditions. In addition, the use of the $\epsilon$-greedy algorithm helps LIM2 dynamically adapt the TTT and hysteresis considering the characteristics of the environment. Overall, LIM2 provides a smart mechanism to handle very high mobility by intelligently selecting the target cell along with the TTT and hysteresis levels. Through simulations, it is noted that LIM2 can significantly improve the handover execution in 5G, leading to smart handling of high speed mobility management in 5G and beyond.  

%, the target cell will have the high signal quality compared to other neighbor cells, and consequently the

\section{Acknowledgement}
This work was supported by the Canada Research Chair Program tier-II entitled ``Towards a Novel and Intelligent Framework for the Next Generations of IoT Networks''.

\bibliographystyle{IEEEtran}
\bibliography{reference}

%---------------------------- Biography ----------------------------------

\begin{IEEEbiography}[{\includegraphics[width=1in,height=1.25in,clip,keepaspectratio]{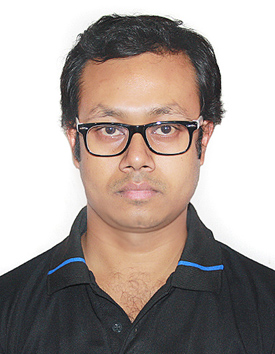}}]{Raja~Karmakar}
Raja Karmakar completed his Bachelor of Technology (B.Tech.) in Computer Science and Engineering from Government College of Engineering and Leather Technology, Kolkata, India and Master of Engineering (M.E.) in Software Engineering from Jadavpur University, Kolkata, India. He received his Doctor of Philosophy (Ph.D.) from Jadavpur University, Kolkata, India. Currently, he is an Associate Professor in the Department of Computer Science and Engineering at Techno International New Town, Kolkata, India. He had done a postdoctoral research at \'{E}cole de Technologie Sup\'{e}rieure (\'{E}TS), Universit\'{e} du Qu\'{e}bec, Montr\'{e}al, Canada. 
    
His research area includes computer systems, wireless networks, mobile computing, IoT, machine learning and UAV communications.   
\end{IEEEbiography}

\begin{IEEEbiography}[{\includegraphics[width=1in,height=1.25in,clip,keepaspectratio]{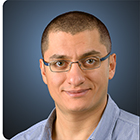}}]{Georges~Kaddoum}
Georges Kaddoum (M'11) received the Bachelors degree in Electrical Engineering from the \'{E}cole Nationale Sup\'{e}rieure de Techniques Avanc\'{e}s (ENSTA Bretagne), Brest, France, and the M.S. degree in Telecommunications and Signal Processing (circuits, systems, and signal processing) from the Universit\'{e} de Bretagne Occidentale and Telecom Bretagne (ENSTB), Brest, in 2005 and the Ph.D. degree (with honors) in Signal Processing and Telecommunications from the National Institute of Applied Sciences (INSA), University of Toulouse, Toulouse, France, in 2009. He is currently an Associate Professor and Tier 2 Canada Research Chair with the \'{E}cole de Technologie Sup\'{e}rieure (\'{E}TS), Universit\'{e} du Qu\'{e}bec, Montr\'{e}al, Canada. In 2014, he was awarded the \'{E}TS Research Chair in physical-layer security for wireless networks. Since 2010, he has been a Scientific Consultant in the field of Space and Wireless Telecommunications for several US and Canadian companies. He has published over 200+ journal and conference papers and has two pending patents. His recent research activities cover mobile communication systems, modulations, security, and space communications and navigation. Dr. Kaddoum received the Best Papers Awards at the 2014 IEEE International Conference on Wireless and Mobile Computing, Networking, Communications (WIMOB), with three coauthors, and at the 2017 IEEE International Symposium on Personal Indoor and Mobile Radio Communications (PIMRC), with four coauthors. Moreover, he received IEEE Transactions on Communications Exemplary Reviewer Award for the year 2015, 2017, 2019. In addition, he received the research excellence award of the Universit\'{e} du Qu\'{e}bec in the year 2018. In the year 2019, he received the research excellence award from the \'{E}TS in recognition of his outstanding research outcomes. Prof. Kaddoum is currently serving as an Associate Editor for IEEE Transactions on Information Forensics and Security, and IEEE Communications Letters.
\end{IEEEbiography}

\begin{IEEEbiography}[{\includegraphics[width=1in,height=1.25in,clip,keepaspectratio]{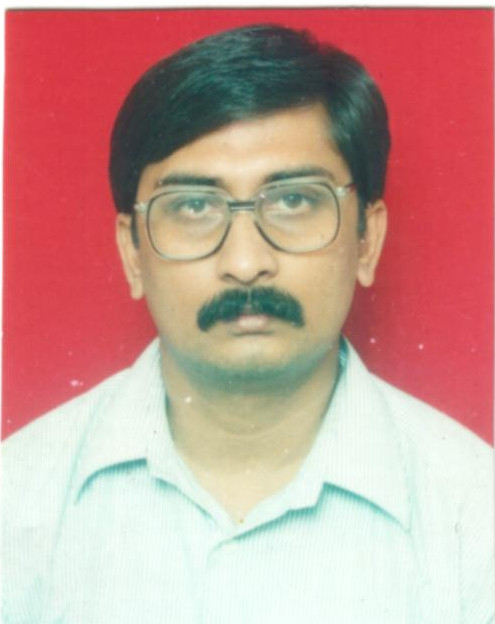}}]{Samiran~Chattopadhyay}
Samiran Chattopadhyay received his Bachelor of Technology (B.Tech.) degree from Indian Institute of Technology, Kharagpur in 1987. He received his Master of Technology (M.Tech.) degree from the same department in 1989 with a Gold Medal for being first in the Institute. He received his Doctor of Philosophy (Ph.D.) Degree from Jadavpur University in 1993. Presently, he is a professor in the Department of Information Technology, Jadavpur University, and Institute for Advancing Intelligence, TCG Centres for Research and Education in Science and Technology, Kolkata, India.  
He has two decades of experience of serving reputed Industry houses such as Computer Associates, Interra Systems India, Agilent, Motorola in the capacity of technical consultant. 

Prof. Chattopadhyay has been working on algorithms for security, bio informatics, distributed and mobile computing, and middleware.
\end{IEEEbiography}

\vfill 

\end{document}